\documentstyle[aps]{revtex}
\chardef\o="1C
\begin{document}
\renewcommand{\theequation}{\arabic{section}.\arabic{equation}}
\title {Theory of hole propagation in one dimensional insulators and
superconductors}
\author{S. Sorella}
\address{
Istituto Nazionale di Fisica della Materia and \\
International School for Advanced Studies, Via Beirut 4, 34013 Trieste, Italy}
\author{A. Parola}
\address{ Istituto Nazionale di Fisica della Materia and \\
Istituto di Scienze Fisiche, Universit\'a di Milano, Via Lucini 3, Como, Italy.}
\maketitle
\begin{abstract}\widetext
The dynamical properties of hole motion in an antiferromagnetic background 
are determined in one dimensional models in zero magnetic
field, where spin isotropy holds, as well as in an external magnetic field. 
The latter case is also
relevant, via particle-hole transformation, to the problem of hole
propagation in one dimensional ``superconductors". The singularities
in the spectral function are investigated by means of bosonization
techniques and perturbation theories. Results are then compared with 
Bethe ansatz solutions and Lanczos diagonalizations. 
The formalism also leads to
interesting connections to the single impurity problem in Luttinger liquids.
A rich structure is found in the spectral function whenever spin isotropy
is broken, suggesting the presence of exotic momentum dependence 
in photoemission spectra of (quasi) one dimensional materials.
\end{abstract}
\pacs{75.10.Jm,71.27+a,74.25.Gz}

\setcounter{equation}{0}

\section{Introduction}
The spectral properties of a single hole in a quantum
antiferromagnet still represent an outstanding problem in the
physics of strongly correlated electron systems. Although more
than 25 years elapsed since the seminal work of Brinkman and Rice
\cite{br} (BR) there is still no consensus on the nature of hole motion
(coherent or incoherent) or on the features of the long range distortion 
induced by the hole on the antiferromagnetic ordering. On the other hand, a
full understanding of the dynamics of a single hole is clearly 
required before the problem of the hole-hole effective interaction, 
mediated by the magnetic background, can be addressed. This issue, relevant 
in the low doping regime, is a key problem in the framework of 
high temperature superconductivity. Furthermore, 
recent developments in angle resolved photoemission and inverse photoemission 
experiments\cite{wells,aebi,quasi1d}, have made possible to extract 
the momentum dependent spectral function in several compounds, including 
high temperature superconducting materials at stoichiometric composition,
which are good quantum antiferromagnets. The photoemitted electron leaves a 
mobile hole in the spin background: therefore these experimental studies 
directly address the problem of hole propagation in systems where electron 
correlations play a key role.

A widely accepted model to describe the physics of a quantum antiferromagnet is
the well known Hubbard model \cite{review} at half filling 
(one electron per site):
\begin{equation} 
\hat H=-t \sum\limits_{<i,j>} (c^{\dagger}_{i \sigma} c_{j \sigma} + 
{\rm h.c. } )
+ U \sum_i (n_{i \uparrow}- {1\over 2}) ( n_{i\downarrow}-{1\over 2}) 
\label{hubbard}
\end{equation}
where $c^{\dagger}_{i \sigma}$ ($c_{i\sigma}$) is the  creation
(annihilation) operator  of an electron with spin $\sigma$ at the lattice site
$i$ and the symbol  $<i,j>$ indicates nearest neighbor summations over an
hypercubic bipartite lattice in arbitrary spatial dimension $d$. Henceforth 
even (odd) values for $i$ indicate  conventionally one of the two sublattices.
The operator $n_{i \sigma}=c^{\dagger}_{i \sigma} c_{i \sigma}$ 
is the  number operator of a  particle with spin $\sigma$ at the
given site $i$. This model is defined in a finite lattice with $L$ sites 
and standard periodic boundary conditions. 
When  the total number of particles $N_c=N_\uparrow+N_\downarrow$  
equals the number of sites, (i.e. at $\rho=N_c/L=1$) this Hamiltonian is 
believed to develop a gap in the charge excitation spectrum. 
This is actually rigorous in one dimension where the exact Lieb and Wu 
solution \cite{liebwu} yields a finite gap for arbitrary repulsion $U>0$ 
and magnetization per site $\mu={N_\uparrow-N_{\downarrow} \over 2 L }$. 
A particle hole transformation 
\begin{equation}
c^{\dagger}_{i \uparrow} \to (-1)^i c_{i \uparrow}
\label{phole}
\end{equation}
maps the half filled Hubbard model at $U>0$ and magnetization 
$\mu$ into the same model at $\mu'=0$, negative interaction ($U'=-U$) and
density $\rho'=1 - 2 \mu$. This mapping also shows that the problem of a 
single hole in an antiferromagnet at non zero magnetic field is relevant to 
understand the photoemission spectra in superconductors: in fact, there is 
quite a robust numerical evidence\cite{scalapino} that the negative $U$ model 
is an $s$-wave superconductor in $d=2$, and in one dimension the exact Bethe
ansatz solution predicts  quasi-long range order in the ground 
state \cite{review1d}.

Although the treatment of a single hole ($N_c=L-1$)  might seem a major
simplification,  there are only few results valid and accepted in more than one
dimension. The Nagaoka theorem is a remarkable exception, stating  
that for $d >1$ the ferromagnetic state with maximum total spin 
$S= {1 \over 2}(L-1)$ 
is the unique ground state of the infinite $U$ Hubbard model, apart for the 
trivial degeneracy of the  $2 S+1 $ spin components \cite{nagaoka}.

At strong coupling the Hubbard model at half filling is mapped into a standard
Heisenberg model with antiferromagnetic superexchange $J= { 4 t^2 \over U}$. 
The presence of a single hole modifies slightly this mapping: each site is
singly occupied and the hole hops from site to site weakening 
local antiferromagnetic correlations. This process is described by
the so called $t-J$ model:
\begin{equation}\label{tj}
\hat H=-t \sum_{<i,j>,\sigma} (c^\dagger _{i\sigma}c_{j\sigma} + h.c. ) +
J \sum_{<i,j>} ({\vec S}_i \cdot {\vec S}_j - {1 \over 4} n_i n_j ).
\end{equation}
where the constraint of no double occupancy is understood.
The exact mapping from Hubbard to $t-J$ also includes a three site 
term, which is neglected here because is believed not to change the
physics of the model, at least in one dimension \cite{ogata,parola}.
The last, density dependent, contribution in Eq. (\ref{tj}) 
can be also dropped for our purposes because it is effective 
only when more than one hole is present in the system.

This work extends and develops the analysis presented in
a previous Letter \cite{letter} about the spectral properties of 
hole motion in {\sl one dimensional} models of correlated electrons.
In such a case, spin charge decoupling allows to describe the low energy 
physics by an effective Hamiltonian $\hat H$ written as the sum
of two  commuting parts $\hat H=\hat H_\sigma+\hat H_\rho$, 
the former governing the
spin degrees of freedom (spinons) and the latter the charge ones (holons).  

This picture is by no means new in the field of one dimensional 
electron systems: standard analytical treatments show that, if the
excitation spectrum of the model is gapless, the low energy physics 
in both charge and spin sectors is described by a Luttinger liquid model 
\cite{solyom,haldane,schulz} for generic microscopic Hamiltonians.
The case is different in the Hubbard model for positive $U$ at half 
filling or, for negative $U$, at zero magnetic field and 
arbitrary density: Only
one of the two sectors is gapless and the renormalization group (RG) 
equations of the generic (the so called $g-$ology) model flow 
to strong coupling. The $g-$ology model parameterizes the most general 
low energy interaction present in one dimensional, translationally invariant
systems of spin one-half electrons.  The model depends on several coupling
constants $g_i$, with $i=1,\dots 4$, which may also have a spin dependence 
$\parallel$ and $\perp$ for electrons interacting with the same or with 
opposite spins, respectively \cite{solyom}. Conventionally,
$g_1$ refers to backward scattering, $g_2$ and $g_4$ to forward scattering
and $g_3$ to Umklapp scattering, the latter present only in lattice models
at commensurate fillings. When the coupling constants $g$'s go to strong 
coupling under the RG flow, they should cross the exactly solvable Luther-Emery 
line leading to a spin gap for $ g_{1 \parallel} < |g_{1\perp}|$  and a 
charge gap for $-2 g_2 < |g_3| $. The Hubbard model at half
filling belongs to the first class at negative $U$ and to the second at
positive $U$, thereby providing a model Hamiltonian which encompasses 
the most general strong coupling fixed points. Therefore,
understanding the Hubbard model at half filling and arbitrary
magnetization would shed light on the physical behavior at the Luther-Emery
fixed point. This would be particularly valuable for the
dynamical properties because the retarded Green function 
\begin{equation} \label{green}
G_\sigma(p,\omega)= -\,<\Psi| c^{\dagger}_{p \sigma}\, (\omega-\hat H + E_0 +
i\,\eta\, )^{-1} c_{p \sigma} |\Psi>
\end{equation}
of the Luther-Emery model is not exactly known, although a widespread 
prejudice ascribes no interesting features to this correlation 
function \cite{review1d,korepin} due to the presence of a gap in
the excitation spectrum.
 
A first study of the single hole Green function in quantum antiferromagnets
(QAF) was performed by Brinkman and Rice \cite{br} who calculated 
the Green function  of the $t-J$ model by neglecting quantum fluctuations, 
i.e. by replacing  the spin exchange interaction by its Ising form:
$\vec S_i \cdot \vec S_j \to S^z_i S^z_j$ in Eq.(\ref{tj}).  In the absence 
of holes, the N\'eel state is the classical ground state of the model, 
whereas the lowest spin excitation has a gap $\sim J$. This Hamiltonian 
can be thought of as the strong coupling limit of a model characterized by 
a gap both in the spin and in the charge sectors (say the half filled 
Hubbard model with uniaxial spin anisotropy). The problem of hole motion
in this system has been studied in any dimensions, but its solution in 
$d=1$ is particularly simple and instructive. When the hole hops in
the lattice it leaves a defect in the N\'eel background. As a result, 
the spectral weight $A(\omega,p) = { 1\over \pi} \, {\rm Im}\,G(p,\omega)$
shows a delta function contribution at the lowest excitation energy with non 
zero quasiparticle weight $Z$, together with an incoherent band separated 
by a gap. A brief discussion of these results is contained in 
Section \ref{isingsec}. The situation is similar in the limit of infinite 
dimensions \cite{vollhardt} where there is no incoherent contribution and only 
a series of $\delta$ function peaks at higher energies is left. The inclusion 
of quantum fluctuations, however, drastically changes this simple picture due 
to the presence of gapless excitations in the magnetic background. This is
the subject of the present study which we organized as follows.

In Section \ref{effsec} an extremely useful mapping between the 
one hole Hamiltonian and an effective spin problem is given in some
detail. In Section \ref{impurasec} we discuss an interesting relationship,
which emerges from the previous formulation,
between hole dynamics and the impurity problem in Luttinger liquids.
The main  result of the present work can be summarized in the general 
structure we find in the Green function of all the models we have examined:
\begin{equation}\label{spincharge}
G(p,\tau)=\int{dQ\over 2\pi} G_h(p+Q,\tau)\,Z_p(Q,\tau)
\end{equation}
where $G_h(k,t)$ is just a free propagator for the holon:
${\rm Im} \,G_h(k,\omega) =\pi\,\delta(\omega -\epsilon_h(k) )$,
$\epsilon_h$ being the holon dispersion energy.
The function $Z(Q,\tau)$ is completely determined by the spinon gapless 
excitations, and is highly non trivial with momentum dependent power law 
singularities and  branch cuts. This decomposition is introduced in
Section \ref{fieldsec} which also contains a discussion of the
analytical properties of the function $Z(Q,\tau)$.
Few specific examples, i.e. $t-J_{XY}$ model
and the Bethe ansatz soluble models, are presented in Sections
\ref{txysec} and \ref{hubtjsec} respectively. A numerical
evaluation of the non universal features of the spectral function in the
Hubbard model is carried out in Section \ref{hubcin} while
some conclusions are drawn in Section \ref{endsec}.

\setcounter{equation}{0}

\section{Hole motion in Ising antiferromagnets}
\label{isingsec}

The problem of hole motion in an Ising antiferromagnet can be solved
exactly and provides the simplest model of hole dynamics in a magnetic 
environment
characterized by a {\sl discrete} excitation spectrum. This feature
turns out to be responsible for the different behavior between the Ising
and the XY case which will be discussed in Section \ref{txysec}.
The Hamiltonian describing hole hopping in an Ising model is a simple
generalization of the $t-J$ model where only the $S^z$ component of the
spins is retained \cite{review}:
\begin{equation}
\hat H=-t\,\sum_{i\sigma} \left[ c^{\dagger}_{i\sigma}c_{i+1\sigma}
+c^{\dagger}_{i+1\sigma}c_{i\sigma}\right ] +J\sum_i 
S_i^z S_{i+1}^z 
\label{tising}
\end{equation}
with $J>0$. We consider a chain with an even number $L$ of sites and 
periodic boundary conditions. 
At half filling the ground state is a classical N\'eel 
state $|{\cal N}>$. Out of the
two states obtained by translation of one lattice vector we select the
one with a spin down at the origin. The annihilation of the spin down
electron at the origin defines our starting state 
$|0>=c_{0 \downarrow}|{\cal N}>$.
The Hamiltonian (\ref{tising}) acting on $|0>$ generates states 
which, in the thermodynamic limit, can be uniquely labeled by the 
position of the hole $|R>$. In fact, for $L\to \infty$, closed paths of the 
hole along the ring can be neglected and the retraceable path approximation 
becomes exact \cite{br}. Then, it is easy to check that 
\begin{equation}
\hat H \,|R>=-t\,\left [\, |R-1>+|R+1>\,\right ]+\left 
(E_{\cal N}+{J\over 2}\right ) 
\,|R>-{J\over 2}\,\delta_{R,0} \, |R>
\label{hising}
\end{equation}
where $E_{\cal N}$ is the energy of the N\'eel state. 
The corresponding eigenvalue equation can be easily solved in this subspace:  
In the thermodynamic limit, the energy spectrum consists of a bound 
state with energy $E_b=E_{\cal N}+{1\over 2}J -{1\over 2}\sqrt{J^2+16t^2}$ 
and exponentially localized wavefunction $\psi_b(r)$. Above the bound state
lies a continuum with energies labeled by the wavevector $q\in (0,\pi)$: 
$E_q=E_{\cal N}+{1\over 2} J-2t\cos q$. For every $q$ there are two degenerate
wavefunctions $\psi_q^{\pm}$ classified according to their parity. Odd states
are unaffected by the perturbation $\delta_{R,0}$ in Eq. (\ref{hising})
because the wavefunction vanishes on site while even states, which include the
bound state, have an on site probability given by:
\begin{eqnarray}
|\psi_b(0)|^2 &=& {J\over \sqrt{J^2+16t^2}} \nonumber\\
|\psi_q^+(0)|^2 &=& {32t^2\sin^2 q\over L \left [J^2 + 16t^2\sin^2q\right ]}
\label{psio}
\end{eqnarray}
The spectral function easily follows from the aforementioned properties of 
the eigenfunctions. In fact, the Lehmann decomposition of $A(p,\omega)$
gives: 
\begin{equation}
A(p,\omega)=\sum_s |<s,-p\,|\, c_{p\downarrow}\,|\,{\cal N}>|^2
\delta(\omega-E_s)
\label{lehmis}
\end{equation}
where the sum is over all the one hole states $|s,-p>$ of momentum $-p$
(modulo $\pi$ due to the doubling of the cell induced by the
antiferromagnetic ordering in $|{\cal N}>$).
These states can be quite generally written as Bloch superpositions
of eigenstates $|s>$ of the Hamiltonian (\ref{tising}):
\begin{equation}
|s,-p>={1\over \sqrt{L}} \sum_R e^{ipR} T^R |s>
\end{equation}
where $T^R$ is the $R$ sites translation operator. This identity 
readily gives the matrix element appearing in Eq. (\ref{lehmis})
in terms of the on site value of the eigenfunction $\psi_s(r)$ 
corresponding to $|s>$: $<s,-p| c_{p\downarrow}|
{\cal N}> = {1\over 2}\psi_s(0)$.
Therefore both the energy levels and the matrix elements are independent 
of $p$. This is due to the absence of fluctuations in the classical 
N\'eel state which leads to a local Green function, as already noticed by
Brinkman and Rice \cite{br} in the $J\to 0$ limit of this problem. By
substituting this result into Eq. (\ref{lehmis}) and taking into account
the form of the energy spectrum we find that, in the thermodynamic limit, 
the bound state is singled out because of its finite value of the
on site probability and gives rise to a delta function peak. Instead
the other states merge into an incoherent band:
\begin{equation}
A(p,\omega)={J\over 2\sqrt{J^2+16t^2}}\,\delta(\omega+{\textstyle{1\over 2}}
\sqrt{J^2+16t^2})+\,{2\over \pi} \,{\sqrt{4t^2-\omega^2}\over J^2+16t^2-
4\omega^2}\,\Theta(4t^2-\omega^2)
\label{isingspec}
\end{equation}
where $\omega$ is now measured from the reference value $E_{\cal N} +{1\over 2} J$
and $\Theta(x)$ is the step function.
The incoherent portion has several interesting properties: 
it is a regular, even function of $\omega$ which vanishes at band edges
$\omega=\pm 2t$ for every nonzero value of $J$. At large $J$ ($J> 4t$)
it shows a broad maximum at $\omega=0$ while for $J < 4t$ it has a minimum
at $\omega=0$ and two symmetrical maxima appear at $\omega=\pm {1\over 2}
\sqrt{16t^2-J^2}$. In the $J\to 0$ limit the incoherent part develops 
square root singularities at band edges in agreement with the BR analysis.
The shape of the spectral function for two representative values of the 
coupling $J$ is shown in Fig. 1. 

In conclusion: The exact Green function of this problem is purely local 
and then the hole does not propagate in the Ising antiferromagnet. 
This is conventionally
understood on the basis of the ``string" defect that the hole creates 
in the antiferromagnetic ordering when it hops \cite{br}. However, the
quasiparticle weight $Z$ is finite at all non zero values of $J$:
The hole behaves as a free particle of infinite mass. This is due to
the nature of the excitation spectrum of the Ising model
which does not allow for gapless modes. In fact we will show that
both features of $A(p,\omega)$ will be strongly modified in more
realistic models of hole dynamics.

\setcounter{equation}{0}

\section{The effective hole Hamiltonian}
\label{effsec}
In the following we consider the $t-J$ Hamiltonian (\ref{tj}) defined on 
$L$ sites with periodic boundary conditions. Our task is to derive an effective
spin Hamiltonian describing how the hole hopping processes perturb
the antiferromagnetic background in the particular case of single hole doping.
In classical physics this would correspond to a Galileo transformation
from the laboratory frame to the reference frame locally at rest with 
respect to the hole. This transformation can be easily generalized to
quantum mechanics \cite{zhong} and the derivation can be performed 
in arbitrary dimension. As a first step we notice that the $t-J$ Hamiltonian 
(\ref{tj}) is translationally invariant and then any one hole state with 
definite momentum $k$ and spin $\uparrow$ can be written as:
\begin{equation} \label{state}
|\psi_k> = {1\over \sqrt{L} }   \sum_{R=0}^{L-1}   e^{-i k R} 
T_L^R c_{0 \downarrow} |\sigma_0>  
\end{equation}
where $|\sigma_0>$ is a {\sl suitable} spin state with the spin at the origin 
$R=0$ fixed to $\downarrow$ and $T_L$ is the spin translation operator defined
by the transformation property:
\begin{equation}
T_L \vec S_R  T_L^{-1} = \vec S_{R+1}.
\label{deftra}
\end{equation}
where periodic boundary conditions (PBC) over the $L$ sites are assumed 
in order to define the effect of translation at the rightmost site. 
The latter relation determines the unitary operator $T_L$ only up to 
an arbitrary phase factor. In order to fix the phase of $T_L$
it is enough to specify the action of the operator on a reference (``vacuum" )
state $|F>$. Here we follow the convention to impose that the 
{\sl ferromagnetic state} $|F>$ is translationally invariant: $T_L |F>=|F>$.  
By substituting the representation (\ref{state}) of the  one hole state into
the eigenvalue equation for the $t-J$ Hamiltonian we find that $|\psi_k>$ 
is an exact eigenstate of the $t-J$ Hamiltonian if and only 
if $|\sigma_0>$ is an eigenstate of the following effective {\sl spin} model:
\begin{equation} \label{hamilt}
H_k=t\,\left [ e^{-i k} T_\ell +\, h.c.\,\right ] \,+\,
J \left[ \sum_{R=1}^{\ell-1} \vec S_R \cdot 
\vec S_{R+1} \right] 
\end{equation}  
where the hole-translation operator $T_\ell$  is defined  exactly as  $T_L$
(\ref{deftra})  but with PBC on a squeezed chain of $\ell=L-1$ sites,
without the origin $R=0$ where the hole sits. $ H_k$ is indeed defined on 
$\ell$ sites since the hole at the origin is decoupled from the other 
sites, namely $H_k$ commutes with the spin operator 
$\vec S_0$ at the origin. In the following we take the convention to 
set the spin at the origin with down orientation: $S^z_0=-{1\over 2}$.   
Notice that the magnetic part, proportional to $J$, represents a Heisenberg 
model with {\sl open} boundary condition, since all the magnetic bonds 
connecting the spins with the hole are obviously suppressed. Instead the 
hole kinetic term $\hat K$ (i.e. the first term in $H_k$) is written 
in terms of
the translation operator $T_\ell$ which enforces periodic boundary condition
on the squeezed chain.
In this way we effectively traced out, with no approximations, the charge degree
of freedom reducing the one hole problem to a purely, non translationally 
invariant, spin model. The effective Hamiltonian $H_k$ explicitly depends 
on the momentum $k$ of the state showing that the distortion of the 
antiferromagnetic ordering does depend in a non trivial (and non local) way 
on hole motion. 

Also the hole dynamics can be conveniently expressed
in terms of the eigenstates of $H_k$. The spectral function
of a $\downarrow$ hole is written, in Lehmann representation, as
\begin{equation}
A(p,\omega)={1\over \pi} {\rm Im}\,G(p,\omega)= \sum_s \,
|\,<s\,|\,c_{p\downarrow}\,|\,\Psi>\,|^2\,\delta(\omega-E_s+E_0)
\label{spec1}
\end{equation}
where $|\,\Psi>$ is the ground state of the model with no holes and
$|\,s>$ represents a complete set of one hole intermediate states.
The corresponding energies are respectively $E_0$ and $E_s$ while the
momentum space annihilation operator is defined by
\begin{equation}
c_{p\downarrow}={1\over\sqrt{L}}\sum_R\,e^{ipR}\,c_{R\downarrow}
\end{equation}
Note that with the adopted definitions, the Heisenberg ground state
$|\,\Psi>$
has total momentum $N\pi$, where $N$ is the number of spins up in the
squeezed chain $N=S_z+\ell/2$, and $S_z$ is the $z-$component of the total spin.
As a consequence, the intermediate states $|\,s>$ must have momentum $N\pi-p$.
By using the general representation (\ref{state}) of one hole states 
with momentum $k$, Eq. (\ref{spec1}) becomes
\begin{equation}
A(p,\omega)=\sum_s \,|\,<\sigma_s\,|\,n_{0\downarrow}\,|\,\Psi>\,|^2
\,\delta(\omega-E_s+E_0)
\label{spectre}
\end{equation}
where now the sum runs over all the eigenstates $|\,\sigma_s>$ 
of $H_k$ with $k=N\pi-p$. 
This expression shows that the quasiparticle weight of the hole 
is simply expressed as the modulus square of the overlap between
the Heisenberg ground state and eigenstates of the effective 
hole Hamiltonian.

Equation (\ref{hamilt}), specialized to the case of a hole of momentum $p$,
can be equivalently written in
a form which makes explicit connection with the problem of a Heisenberg
model with a {\sl local} perturbation. In fact, by adding and subtracting 
the additional bond operator $\hat J =J \vec S_1 \cdot \vec S_\ell$,
the magnetic part of the Hamiltonian $\hat H_\sigma$ can be made 
translationally invariant (on the {\sl squeezed} chain):
\begin{eqnarray}
\hat H_p&=& \hat K + \hat H_\sigma - \hat J \nonumber\\
&=& 
t\,(-1)^N\,\left [ e^{i p} T_\ell +\,h.c.\,\right ] \, +\,
J \left[ \sum_{i=1}^{\ell} \vec S_i \cdot  \vec S_{i+1} \right] 
-J  \, S_1 \cdot \vec S_\ell
\label{hterm}
\end{eqnarray}
The only extensive, i.e. $O(L)$, term in the total hole Hamiltonian $\hat
H_p$ is $\hat H_\sigma$ which coincides with the usual 
Heisenberg model, while the other two
contributions induce $O(1)$ corrections to the total energy. 
Therefore, the physics of the {\sl bulk} of the spin system is not affected 
by the presence of the hole which, at most, acts as a {\sl boundary term}
on the Heisenberg antiferromagnet. This can be simply understood in two
limiting cases: for a static hole, since $t=0$ the Hamiltonian 
$\hat H_p$ becomes a Heisenberg model with {\sl open} boundary conditions,
while, for $J\to 0$, the last term $\hat J$ is irrelevant and the 
eigenstates of $\hat H_p$ are those of a Heisenberg model with {\sl periodic} 
boundary conditions on a $\ell$ site chain.
A close connection between the presence of the
hole and a change in boundary conditions on the magnetic Hamiltonian
thus emerges quite naturally, in every dimensions, within this formalism.

In one dimension we can proceed further by mapping this spin ${1\over 2}$ model
into a spinless fermion Hamiltonian via standard Jordan-Wigner transformation
\cite{review1d}.
It is convenient to introduce an additional phase factor in the fermion 
creation operators $\psi^\dagger_n$ in order to obtain the usual sign of
the kinetic energy term:
\begin{equation} 
S^+_n=(-1)^n \psi^\dagger_n 
\exp\left [i \pi \sum\limits_{j=1}^n \psi^{\dagger}_j \psi_j \right ]
\label{wigner}
\end{equation}
The magnetic part $\hat H_\sigma$ of the Hamiltonian (\ref{hterm}) becomes:
\begin{equation} 
\hat H_\sigma =
-{J \over 2} \sum_{i=1}^{\ell}( \psi^{\dagger}_i \psi_{i+1} + h.c. )
+ J \sum\limits_{i=1}^{\ell} ({1\over 2} - \psi^{\dagger}_i \psi_i) 
({1\over 2} - \psi^{\dagger}_{i+1} \psi_{i+1})
\label{heff1}
\end{equation}
where the boundary conditions in the first term are periodic  
or antiperiodic for even or odd values of $L+N$, i.e.
$\psi_{\ell+1}\equiv (-1)^{N+L}\psi_1$. As usual, $N$ is the number 
of fermions, which is related to the total magnetization of the original 
spin model by $S_z= N-{\ell\over 2}$. Analogously, the bond term $\hat J$ 
is written as:
\begin{equation}
\hat J= -{J \over 2}\,(-1)^{N+L}\,( \psi^{\dagger}_1 \psi_{\ell} + h.c. ) + J 
({1\over 2} - \psi^{\dagger}_1 \psi_1) 
({1\over 2} - \psi^{\dagger}_{\ell} \psi_{\ell})
\end{equation}
In order to express the hole kinetic term $\hat K$ in terms of
the spinless fermion operators we have to relate the spin translation 
operator $T_\ell$ to the usual fermion translation operator $T_f$
which leaves invariant the fermionic vacuum state and satisfies
$T_f \psi_i T_f^{\dagger}  = \psi_{i+1}$.
Keeping track of the phase factors in the definitions we have
$T_\ell=(-1)^N T_f$. The fermion translation operator is then conveniently 
expressed in terms of the Fourier transformed operators
$\psi_k =\ell^{-1/2} \sum\limits_{j=1,\ell} \psi_j e^{i k j} $
where the momenta $k$ are quantized according to the choice of boundary
condition:
\begin{equation}
T_\ell=(-1)^N T_f =\exp\left [
 i \sum\limits_k (k+\pi) \psi^{\dagger}_k \psi_k \right ]
\label{heff2}
\end{equation}
By use of Eqs. (\ref{heff1},\ref{heff2}) we can express the Hamiltonian 
of a spin down hole of momentum $p$ in the $t-J$ model, 
$\hat H_p$ (\ref{hterm}), in terms of spinless fermion operators:
\begin{equation}
\hat H_p=t\,\exp\left [\,ip+i\sum\limits_{k} k \psi^{\dagger}_k \psi_k
\,\right ]+h.c.
-{J \over 2} \sum_{i=1}^{\ell-1}( \psi^{\dagger}_i \psi_{i+1} + h.c. )
+ J \sum\limits_{i=1}^{\ell-1} ({1\over 2} - \psi^{\dagger}_i \psi_i) 
({1\over 2} - \psi^{\dagger}_{i+1} \psi_{i+1})
\label{htot}
\end{equation}
This concludes our formal manipulations on the 
original problem. We now have an interacting fermion system which can
be studied by means of the powerful techniques developed in the framework
of one dimensional physics, ranging from renormalization group approaches 
to bosonization methods \cite{solyom}. However, before addressing these
issues, it is instructive to dwell on the similarities between the problem of 
hole motion and the effects of local perturbations in Luttinger liquids,
which emerge naturally from the structure of the effective Hamiltonian 
(\ref{hterm}).

\setcounter{equation}{0}

\section{Relationship to the impurity problem}
\label{impurasec}

In this Section we like to investigate the close relationship 
between the physics of a single hole in 
the $t-J$ model and the impurity problem in Luttinger liquids. 
The connection between these two different problems becomes apparent 
when we take advantage of the previously discussed Galileo transformation. 
The Hamiltonian $\hat H_p$ governing the dynamics of the hole represents a 
Heisenberg model with open boundary conditions plus the hole kinetic 
contribution which involves the translation operator $T_\ell$. 
Furthermore, note that the matrix
elements appearing in the spectral function (\ref{spectre}) are
related, through Eq. (\ref{state}), to the overlaps between the ground
state $|\,G>$ of the unperturbed Hamiltonian ($\hat J=\hat K=0$) on
$L$ sites and that of the perturbed one ($\hat J\ne 0$) in the
squeezed chain of $\ell$ sites. The difference in the number of sites
of the chain defining the perturbed state can also be thought of as the
local perturbation induced by the removal of the two bonds connecting
the origin in the $L$ site ring.

Let us focus our attention on the Hamiltonian of a hole of momentum $p$ in 
the $L$-site $t-J$ model with $XY$ spin anisotropy: the $t-J_{XY}$ chain.
In fermion representation $\hat H_p$ is:
\begin{equation}
\hat H_p =t\,\exp \left [ip+i\sum\limits_{k} 
k \psi^{\dagger}_k \psi_k\right ] + h.c. 
-{J\over 2} \sum_{i=1}^{\ell} \left [\psi^{\dagger}_i \psi_{i+1} + h.c.\right ]
+{J^\prime \over 2} \,(-1)^{N+L} 
\left [\psi^{\dagger}_{\ell} \psi_{1} + h.c.\right ]
\label{wigj}
\end{equation}
where $J^\prime =J$ represents the perturbation induced by the hole.
In order to be specific, we consider the model with $N=2\nu+1$ 
fermions (spinons)
in a lattice of $L-1=\ell$ sites. The kinetic operator $\hat K$ commutes with
the magnetic term in $\hat H_\sigma$ and therefore, at 
$J^\prime =0$, the eigenvectors 
of $\hat H_p$ coincide with those of the $XY$ model on $\ell$ sites 
with PBC while the eigenvalues are shifted by $2\,t\,\cos(p+Q)$ where $Q$ 
is the momentum of the spin state. The physics of hole motion is contained in
the perturbation term proportional to $J^\prime $, therefore it is convenient to
study the effects of the local perturbation as a function of 
its strength $J^\prime $. In the static limit ($t=0$) the hole kinetic term 
$\hat  K$ is suppressed and we recover the Hamiltonian of a $XY$ model with a 
weak bond.
This problem has been extensively studied in the past \cite{affleck,kane}:
the last term in Eq. (\ref{wigj}) 
is a relevant perturbation and at long wavelengths 
the system behaves as a chain with a missing bond. As a consequence,
the overlap $\zeta$ between the state with $J^\prime =0$ and the state 
with finite $J^\prime $ goes to zero in the thermodynamic limit with 
a universal exponent $X$ independent of the strength of the perturbation:
\begin{equation}
\zeta= < J^\prime \, | \, G > \,\propto \,L^{-X}
\label{zetaxy}
\end{equation}
According to conformal field theory, the exact value of the exponent $X$ 
can be found analytically by taking the difference 
between the size corrections of the open and the periodic chain and dividing by
$2\pi v_s$, where $v_s$ is the spinon Fermi velocity.
In the XY model we get $X=\mu^2/4 + 1/16$ where $\mu\in (-1/2,1/2)$ is
the magnetization of the XY model along the $z$ direction.

The recoil of the hole, embodied in the hole kinetic term $\hat K$, 
qualitatively changes this picture as we already anticipated.
The perturbation $J^\prime $ is now marginal: it does not drive the model 
towards open boundary conditions but instead it changes the boundary conditions 
of the fermionic model in Eq. (\ref{wigj}) introducing phase shifts for right 
and left moving spinons which depend on the strength of the perturbation
$J^\prime $ and vanish as $J^\prime \to 0$.

The effect of the kinetic term can be understood by use of first order 
perturbation theory in the parameter $J^\prime $ in Eq. (\ref{wigj}). 
For $J^\prime =0$ the ground state $|G>$ 
is non degenerate and can be represented by a Slater determinant of plane 
waves with PBC. At first order, the ground state becomes:
\begin{equation}
|J^\prime >=
|G>+{J^\prime \over 2 \ell}\sum_{k\ne q}\left( e^{ik}+e^{-iq} \right )\,
{<k,q|\psi^{\dagger}_k \psi_{q} |G>\over \epsilon_{k,q}}\, |k,q>
\label{pert}
\end{equation}
where the sum runs over the momenta of particle-hole excitations
of the unperturbed system and $\epsilon_{k,q}$ are the corresponding 
excitation energies: 
\begin{equation}
\epsilon_{k,q} = J \left [\cos q - \cos k\right ] +
2t\left [\cos(p+k-q)- \cos p \right ]
\label{excita}
\end{equation}
The matrix element in Eq. (\ref{pert})
$<k,q|\psi^{\dagger}_k \psi_{q} |G>$
is unity provided $q$ belongs to the Fermi sea while
$k$ lies outside the Fermi surface. Perturbation theory fails in
the thermodynamic limit due to low energy excitations which may be of
``forward" (i.e. $q\sim \pm k_F$ and $k\sim \pm k_F$) or ``backward"
type (i.e. $q\sim \pm k_F$ and $k\sim \mp k_F$). However, for any
$t\ne 0$ the backward scattering is cut off by the recoil and does
not introduce singularities in perturbation theory, as can be easily 
checked by use of Eq. (\ref{excita}). Instead, forward scattering
is always singular leading to the vanishing of the overlap $\zeta$
(\ref{zetaxy}) in the thermodynamic limit. 

In order to better understand the change in the state induced by the 
boundary term $J^\prime $, let us consider a different problem, namely,
the way the ground state changes due to a weak modification of 
the boundary conditions.
Let us take as unperturbed state a Slater determinant of plane waves
$\phi_q(r)=e^{iqr}\ell^{-1/2}$ with momentum quantization appropriate
to PBC: $\ell q_j=2\pi j$. A change in boundary conditions involves 
the introduction of certain momentum dependent phase shifts $\delta_k$. 
Therefore, we take as perturbed state another Slater determinant of
plane waves $\psi_k(r)$ with generalized momentum quantization rule:
$\ell k_j=2\pi (j+\delta_k)$. Here and in the following, the phase shifts 
will be measured in units of $2\pi$. For weak perturbations, the phase shifts 
will be small and we can keep only $O(\delta_k)$ terms in the
expansion of the state in the unperturbed basis. As usual, to linear 
order, only a single particle-hole excitation $|k,q>$ is allowed and the
weight of the corresponding contribution turns out to be:
\begin{equation}
{\pi \delta_k\over \ell \sin {k-q\over 2}} 
\label{slater}
\end{equation}
for $k\ne q$.
Again, we see that a change in boundary conditions leads to singularities
only for forward scattering as in the hole problem when the effect 
of the recoil is taken into account. The comparison can be made quantitative
by linearizing the momentum dependence about the Fermi points in Eqs.
(\ref{slater},\ref{pert}) and matching the two expressions.
The effective phase shifts induced by the perturbation $J^\prime $
at long wavelength in the Hamiltonian (\ref{wigj}) are then given 
to $O(J^\prime )$ by
\begin{equation}
2\pi \delta_{\pm} = {J^\prime  \cos k_F \over 2t\sin p \mp J\sin k_F}
\label{phasepert}
\end{equation}
at the Fermi points. The two signs refer to the right and
left moving spinons which in general have different phase shifts. Notice 
that the resulting phase shifts depend, in general, on the density as well 
as on the total momentum $p$ of the hole. Following the well known
analysis of Anderson's orthogonality catastrophe \cite{orto}, 
we then find that
the overlap between the states before and after the perturbation
should vanish in the thermodynamic limit with an exponent 
$X=(\delta_+^2+\delta_-^2)/2$. Few details are reported in appendix 
(\ref{overapp}).

In conclusion, we can interpret the effects of the presence of a hole in an 
antiferromagnetic background as a change of boundary conditions in
the corresponding spin problem. If the recoil of the hole is neglected, 
(i.e. $t=0$) both forward and backward scattering between the holon and
the spinons are relevant. In this case, the hole effectively breaks the
spin ring and at low energy the model becomes equivalent to an open
spin chain (in the hole reference frame). Instead, for $t\ne 0$ the
backward scattering channel is cut off and the perturbation induced by
the hole becomes marginal. The presence of the holon introduces, via the
forward scattering channel, phase shifts in the boundary conditions
of the spin chain. Spinons can then propagate {\it through} the site
where the hole sits. 

These results have been obtained analytically for a $XY$ model but,
in view of the universality of the Luttinger Liquid description of
1D correlated systems, we expect that our picture remains valid for general 
$XXZ$ spin chains, including the isotropic Heisenberg point.

\setcounter{equation}{0}

\section{field theoretical analysis}
\label{fieldsec}
In this Section we address the problem defined by the Hamiltonian 
$\hat H_p$ (\ref{htot})
in fermionic representation  by use of the bosonization method.
The bulk term $\hat H_\sigma$ (\ref{heff1}) just represents a Heisenberg model
with periodic boundary conditions which has been
extensively studied in the past \cite{solyom}.
The renormalization group approach applied to such a problem 
shows that both backward scattering and
Umklapp scattering terms are  marginally irrelevant: at long wavelength 
the model is characterized only by the interactions of the forward type
with non universal coupling constants renormalized by the RG flow. 
This fixed point
Hamiltonian represents a Luttinger model which can be exactly mapped into a
free bosonic system following the procedure of  Mattis and Lieb 
\cite{liebmattis}, the bosons representing density fluctuations. 
As suggested by the analysis of Section \ref{impurasec}, the 
additional terms present in $\hat H_p$ slightly modify this picture,
appropriate for the undoped antiferromagnet.
The combined effect of $\hat K$ and $\hat J$ on the long wavelength 
physics is equivalent to a change in the boundary conditions of the 
Luttinger model from periodic to skew:
Although the original problem (\ref{hterm})is 
not translationally invariant due to the bond term $\hat J$, the low energy 
fixed point Hamiltonian develops effective boundary conditions which 
depend on the non universal parameters characterizing the microscopic
model. In particular we expect that the boundary
conditions will continuously evolve from periodic at small $J/t$ to
generally skew at finite $J/t$ with the exception of the zero magnetic 
field case where the boundary conditions at the fixed point remain 
periodic for any $t\ne 0$ due to the presence of the additional $SU(2)$ 
spin symmetry, as will be shown later. The only singular point is at $t=0$
where the hole kinetic contribution vanishes and $\hat H_p$ describes a
Heisenberg model with open boundary conditions. In the following,
we will exploit this picture extracting quantitative predictions
which will be later compared with the exact solution in specific models. 

On this basis we are led to consider a low energy problem defined by a 
long wavelength effective Hamiltonian $\hat H_p$ sum of two 
{\sl commuting} terms: a Luttinger model with skew boundary conditions plus 
the hole kinetic term 
\begin{eqnarray}
\hat K &=&\epsilon_h(k_h)
\label{boseK}\\
k_h &=& p+Q
\label{conserva}
\end{eqnarray}
Here we have defined $k_h$ as the {\sl holon momentum} and $Q$ is the 
{\sl spinon} momentum which obeys the quantization rule {\sl appropriate} 
for the skew boundary conditions of $\hat H_\sigma$. Eq. (\ref{conserva}) 
therefore represents 
momentum conservation of the charge and spin excitations. Generally, we expect 
that the renormalization group flow modifies the effective hole band 
in a non universal way: $2t\cos(k_h)\to \epsilon_h(k_h)$. 

Clearly, such a form of the long wavelength Hamiltonian may describe only
the low energy part of the excitation spectrum of the effective hole 
Hamiltonian (\ref{htot}). Therefore, an elementary spin excitation 
which changes the total spinon momentum $Q$ must be accompanied by
a corresponding variation of the total momentum $p$ so that 
the holon momentum $k_h$ is not modified. Otherwise the
hole kinetic term $\hat K$ would contribute a finite amount to the excitation
energy and a low energy excitation in the spin channel would not correspond 
to a low lying excitation of the total Hamiltonian.
In other words, the low energy spin excitations described 
by the effective spinon Hamiltonian with skew boundary conditions
represent the physical excitation spectrum at fixed holon momentum $k_h$.

The overall picture of the low energy physics of the one hole problem
is confirmed by the exact solution of Bethe Ansatz 
models which also allows for a quantitative analysis of their long 
wavelength properties as will be shown later. 

As a first step, we now generalize the bosonization procedure 
in Luttinger models with skew boundary conditions (Sec.
\ref{bosesec}), then we analyze in some detail the form of the hole 
kinetic operator present in the effective Hamiltonian,
the {\sl possible} boundary conditions which may occur in microscopic models 
and  the corresponding finite size corrections to the ground state 
energy of this effective model (Sec. \ref{boundsec}). Finally we discuss
some implication of the previously obtained results including 
the expected asymptotic form of the hole Green function 
(Sec. \ref{greensec}).

\subsection{The long wavelength Hamiltonian: Bosonization}
\label{bosesec}

At low energy, the important degrees of freedom for a many fermion
system are  those close to the Fermi momenta $k_F^{\pm}$ for the 
left and right movers, i.e.  $k\sim +k_F^+$ in the right ($+$) branch and 
$k\sim -k_F^-$ in the left ($-$) branch. 
As usual, the two branches are extended 
to infinity, within the assumption that the low energy physics is not 
affected by this approximation \cite{solyom}. This extension allows to 
define two fields $\psi_+(x)$ and $\psi_-(x)$ representing, 
in the continuum limit, the annihilation operators for spinless 
fermions on the right and left branch. The continuum limit of the 
original fermionic field $\psi^{\dagger} (x)$ is then given by
the linear combination:
\begin{equation} \label{defrl}
 \psi^{\dag} (x) \propto  e^{i k_F^+ x } \psi^{\dagger}_+ (x) + 
e^{-i k_F^- x } \psi^{\dagger}_- (x)
\end{equation}
where we have kept the distinction between right and left Fermi momenta
$k_F^{\pm}$ which may in principle differ. 
The origin of the fermionic field $x=0$ has been chosen to match with 
the first site $i=1$ of the squeezed chain of length $\ell$. 
As stated before, the fields $\psi_{\pm}(x)$ obey skew
boundary conditions:
\begin{equation}
\psi_{\pm}(x+\ell)=e^{2\pi i \delta_{\pm}} \psi_{\pm}(x)
\label{skew}
\end{equation}
with arbitrary phase shifts $\delta_{\pm}$. 

The Luttinger model is defined 
by right and left moving fermions with kinetic term and interactions of
the forward type. This allows to express the effective low 
energy theory in terms of two bosonic fields defined as bilinear 
combinations of the fermionic operators. More precisely, following
Mattis and Lieb \cite{liebmattis}, the Fourier transform 
$N_{\pm}(q) = \int dx  e^{-i q x }  N_{\pm}(x) $ of the operators
$N_{\pm}(x)= \psi^{\dagger}_{\pm} (x) \psi_{\pm} (x)  + {\rm const.}$ satisfies
non trivial commutation rules
\begin{equation} \label{commrule}
\left[ N_{\pm} (q) ,N_{\pm} (-q)  \right]= \mp {\ell  q \over 2 \pi }
\end{equation}
The above equations suggest the definition of 
the boson field $\Phi(x)$ and its conjugate momentum $\Pi(x)$ via:
\begin{eqnarray}
N_+(x)=\psi^{\dagger}_+(x) \psi_+ (x)- <\psi^{\dagger}_+(x) \psi_+ (x)> 
&=& {1\over \sqrt{4 \pi} }
\left( \Pi(x) + \partial_x  \Phi (x) \right) \nonumber \\
N_-(x)=\psi^{\dagger}_-(x) \psi_- (x) -<\psi^{\dagger}_-(x) \psi_- (x)> 
&=&- {1\over \sqrt{4 \pi} }
\left( \Pi(x) - \partial_x \Phi (x)  \right)
\label{bosrules}
\end{eqnarray}
where the average $<\,>$, taken on the reference ground state of the Luttinger
model, is introduced in order to regularize the divergences.
These operators obey the canonical commutation relations:
\begin{equation}
\left [ \Phi(x), \Pi (x^\prime) \right] = i \delta (x-x^\prime)
\end{equation}
Notice that the densities $N_{\pm}(x)$ satisfy periodic boundary conditions
independently of the phase shifts $\delta_{\pm}$, being bilinear 
combinations of fermionic operators. As a consequence the fields 
$\partial_x\Phi(x)$ and $\Pi(x)$ obey periodic boundary conditions and
the Luttinger Hamiltonian, which can be entirely expressed in terms of the 
bosonic fields ($\Phi(x)$,$\Pi(x)$), has exactly the same gaussian structure 
as for a pure Heisenberg model:
\begin{equation} \label{defh0}
\hat H_\sigma = {v_s \over 2 } \int\limits_0^\ell dx \left[ 
K_\sigma \Pi^2 (x) + {1 \over K_\sigma } ( \partial_x  \Phi  )^2 \right]
+\,{\rm const}
\end{equation}
where $v_s$ and $K_\sigma$ are the renormalized Fermi velocity and the 
dimensionless interaction parameter which characterize the 
long wavelength behavior of the Heisenberg model \cite{review1d}. 

This simple quadratic Hamiltonian can be diagonalized by introducing
normal modes and this procedure leads to the familiar Luttinger Liquid
energy spectrum of the one dimensional Heisenberg model.
However, here we are  interested in the change in the finite size corrections
to the energy spectrum induced by the presence of the hole. Therefore, the
normal modes must be carefully defined in the $\ell$ site chain.
As a first step note that two quantum numbers  labeling 
the eigenstates can be defined: The two operators
\begin{eqnarray}
\Pi^* &=& 
\int\limits_0^\ell \Pi (x) dx = 
\sqrt{\pi}\int\limits_0^\ell \left [ N_+(x)-N_-(x)\right ] dx 
\nonumber \\
\Delta\Phi&=&\left [ \Phi (\ell)- \Phi(0) \right ] = 
\sqrt{\pi}\int\limits_0^\ell \left [ N_+(x) + N_-(x)\right ] dx 
\label{quantumno}
\end{eqnarray}
{\sl commute} each other and also commute with $\hat H_\sigma$,
being related to the total number of fermions on the two branches,
which are conserved quantities in the Luttinger model.
Therefore the diagonalization of the quadratic form (\ref{defh0}) 
can be performed in each sector defined by the pair of quantum numbers
($\Delta\Phi$, $\Pi^*$). In particular, the choice $\Delta\Phi=\Pi^*=0$
identifies the reference state introduced in Eq. (\ref{bosrules}).
The quantization rules directly follow from the
definition (\ref{quantumno}) of these operators. At fixed total number of 
fermions (i.e. at fixed magnetization in the original model) $\Delta\Phi$
is uniquely determined. A low energy excitation in this 
model corresponds to moving a fermion from the left to the right branch. 
This changes the value of $\Pi^*$ by an even multiple of $\sqrt{\pi}$.
On the other hand, by varying the total magnetization by integers,
$\Delta\Phi$ changes by integer multiples of $\sqrt{\pi}$ while the 
model preserves a non degenerate (degenerate) ground state for even
(odd) $\ell$ due to a change in the boundary conditions 
associated with the Jordan-Wigner transformation. Therefore we conclude 
that the possible values of $\Delta\Phi$ and $\Pi^*$ are 
\begin{eqnarray}
\Delta\Phi &=&\sqrt{\pi}(n+\gamma)\nonumber\\
\Pi^* &=& 2\sqrt{\pi}(m+\delta)
\label{nm}
\end{eqnarray}
Here $m$ and $n$ are integers while $\gamma$ and $\delta$ are two non 
universal real quantities (defined mod(1)) which characterize the ground 
state of the model. Their precise value
is determined by the renormalization group flow which connects the microscopic
model to the Luttinger Hamiltonian.  The bosonization procedure alone
does not fix these quantities uniquely, except when additional symmetries
are present in the microscopic model.  

Now we are ready to introduce the normal modes which diagonalize the quadratic
form (\ref{defh0}). It is convenient to define the periodic field
\begin{equation}
\Psi(x)=\Phi(x)-{x\over\ell}\Delta\Phi
\label{period}
\end{equation}
By substituting into (\ref{defh0}) we get:
\begin{equation} \label{defh1}
\hat H_\sigma = {v_s\over 2 } \int\limits_0^\ell dx \left[ 
K_\sigma \Pi^2 (x) + {1 \over K_\sigma } ( \partial_x  \Psi  )^2 \right]
+{v_s\over 2\ell K_\sigma}(\Delta\Phi)^2
\end{equation}
Next we define canonical creation and annihilation boson operators:
\begin{eqnarray}
\Psi_k&=&{1\over\sqrt{\ell}}\int_0^\ell dx \Psi(x)e^{-ikx}=
\sqrt{K_\sigma\over 2|k|} (a^\dagger_k +a_{-k})\nonumber \\
\Pi_k&=&{1\over\sqrt{\ell}}\int_0^\ell dx \Pi(x)e^{ikx}=
i\sqrt{|k|\over 2K_\sigma} (a^\dagger_{-k} -a_{k})
\label{planew}
\end{eqnarray}
where the values of $k\ne 0$ correspond to periodic boundary conditions 
and are quantized in units of ${2\pi\over\ell}$.
Finally we get the low energy Hamiltonian $\hat H_\sigma$ in diagonal form:
\begin{equation}
\hat H_\sigma=E_0+{v_s\over 2\ell}\left({(\Delta\Phi)^2\over K_\sigma}+K_\sigma
(\Pi^*)^2\right ) +v_s\sum_{k\ne 0} |k|a^\dagger_ka_k
\end{equation}
Here $E_0$ is the reference energy of the non degenerate ground state 
of the model with $\gamma=\delta=0$ which is known, from conformal field
theory \cite{review1d}, to depend on the lattice size $\ell$ as:
\begin{equation}
E_0=e_0\ell -v_s{\pi\over 6\ell}
\label{centralpbc}
\end{equation}
if $e_0$ is the Heisenberg ground state energy per site. Notice that whenever 
$\delta$ or $\gamma$ acquire non zero values only the $1/\ell$ 
finite size corrections to the ground state energy (\ref{defh1})
are modified. The quantum numbers $\gamma$ and $\delta$ then 
correspond to energy contributions induced by a change in the boundary
conditions of the model. This remark will be made more precise
in the following.

\subsection{Spinon momentum and skew boundary conditions}
\label{boundsec}

In the previous Section we investigated the bosonization of the
bulk part of the spin Hamiltonian in some detail. At the Luttinger fixed
point the Hamiltonian $\hat H_\sigma$ can be expressed in terms of bosonic
operators in the standard form (\ref{defh0}) independently of the 
presence of skew boundary conditions (\ref{skew}).
However, the energy spectrum depends on two quantum numbers $(\Delta\Phi,
\Pi^*)$ which should be somehow
related to the particular choice of boundary conditions.
In order to make this relationship more transparent, let us analyze the
bosonization form of the momentum operator (i.e. of the translation 
operator) appropriate to the chosen boundary conditions.

The fermionic translation operator is defined by
$T_f\psi(x)T_f^\dagger=\psi(x+1)$. This unitary transformation can be 
split conveniently into two steps: first a conventional translation of
left and right movers  $\psi_{\pm} (x) \to \psi_{\pm} (x+1) $ (\ref{deftras})
and then a rescaling of the fields:
$\psi_+ \to e^{-i k_F^+} \psi_+  $ and $\psi_- \to e^{i  k_F^-} \psi_-$
(\ref{deftra1}), in order to take into account the phase factors in
(\ref{defrl}). The correct form of $T_f$ is then $T_f=T_1 T_2$ where 
\begin{eqnarray}
T_1 &=& \exp\left\{ i \int\limits_0^\ell dx
\left[ k_F^+ \psi^{\dagger}_+ (x) \psi_+(x) -  k_F^-
\psi^{\dagger}_- (x) \psi_- (x) \right] \right \} 
\label{deftra1}  \\
T_2 &=& \exp \left\{-  \int \limits_0^\ell dx
\left[ \psi_+^{\dagger} (x) \partial_x \psi_+ (x)
 + \psi_-^{\dag} (x) \partial_x  \psi_- (x) \right] \right \}
\label{deftras}
\end{eqnarray}
There is no overall constant in the definition of the translation operator
since the expression (\ref{deftras}) acts as the identity on the 
fermionic vacuum state.
Note that $T_1$ has been already bosonized via the (\ref{bosrules}).
The other factor  $T_2$ is just a conventional translation of the
fields $\psi_{\pm}(x)$ and acts on the bilinear forms $N_{\pm}(x)$ as
the {\sl bosonic} translation operator:
$\Pi(x) \to T_2\Pi(x)T_2^{-1}=\Pi(x+1)$ and $\Phi(x) \to \Phi(x+1)$. Therefore
\begin{equation}
T_2=\exp \left \{ {i \over 2} \int\limits_0^\ell dx \left [ 
\Pi(x) \,  \partial_x \Phi + \partial_x \Phi \, \Pi(x) \right ] \right \} 
\label{trasbose}
\end{equation}
From the previous analysis is clear how to represent the translation operator 
$T_f$ within bosonization. By use of the explicit expressions of the 
terms in $T_1$, $T_2$ and Eq. (\ref{bosrules}) we get
\begin{eqnarray}
T_f&=& e^{i \hat P }   \nonumber \\
\hat P &=& Q_0+{k_F^+ + k_F^-  \over 2\sqrt{\pi} } \Pi^*
+ {k_F^+-k_F^- \over 2 \sqrt{\pi}} \Delta\Phi
+ {1 \over 2} \int\limits_0^\ell dx \left [
\Pi(x) \, \partial_x \Phi+ \partial_x  \Phi \, \Pi(x) \right ] \nonumber\\
&=& Q_0 + \left (k_F^+ + k_F^- +2\pi{(n+\gamma)\over\ell}\right) (m+\delta)
+ {k_F^+-k_F^- \over 2 } (n+\gamma)
+ {1 \over 2} \int\limits_0^\ell dx \left [
\Pi(x) \, \partial_x \Psi+ \partial_x  \Psi \, \Pi(x) \right ] 
\label{deftrasla}
\end{eqnarray}
where in the last line we took advantage of  the canonical transformation
(\ref{period}) and we expressed $\Pi^*$ and $\Delta\Phi$ by use of 
Eqs. (\ref{nm}). The additive constant $Q_0$ represents a finite 
contribution which depends on the bandwidth cut-off of the Luttinger model
but not on the chain length $\ell$ and is obtained by the substitution of
Eq. (\ref{bosrules}) into Eq. (\ref{deftra1}).
Its value is determined by requiring that in the ground state 
($n=m=0$) the spinon momentum, in the $\ell \to \infty$ limit, tends 
to $k_F$, i.e. to the momentum of the translationally invariant case without 
perturbation induced by the hole ($\hat J=\hat K=0$ in Eq. \ref{hterm}).
This gives $Q_0=k_F(1-2\delta)$ for odd $\ell$. 
The total momentum operator $\hat P$ commutes with
the bulk Hamiltonian $\hat H_\sigma$ and then each spinon state 
is characterized by a spinon momentum $Q$ which is related to the
total momentum $p$ and to the holon momentum $k_h$ by Eq. (\ref{conserva}).

Through the bosonized form of the fermion translation operator it
is also possible to relate the phase shifts $\delta_{\pm}$ which define
the boundary condition of the microscopic model to the effective
parameters ($\gamma$,$\delta$) entering the low energy Hamiltonian.
The elementary excitation which changes the value of $\Pi^*$ 
corresponds to moving a fermion from the left to the right branch.
In fermionic representation, this excitation carries momentum 
$k_F^-+k_F^++{2\pi\over \ell}(\delta_+-\delta_-)$. Instead, in
bosonic representation, this corresponds to $n=0$ and $m=1$.
In order to match the change in total momentum 
due to such an excitation we have to identify 
\begin{equation}
\gamma=\delta_+-\delta_- \quad {\rm mod}(1)
\label{gam}
\end{equation}
Analogously, a low energy and low momentum excitation which changes 
the total spin corresponds to adding one fermion to the right branch 
and one fermion to the left branch. The change in momentum is then 
$k_F^+-k_F^-+{2\pi\over \ell}(\delta_++\delta_-)$ while the quantum
numbers are $n=2$ and $m=0$. Matching the two expressions gives:
\begin{equation}
\delta={\delta_++\delta_-\over 2} \quad {\rm mod}(1)
\label{delt}
\end{equation}

This concludes the bosonization of the long wavelength effective 
Hamiltonian, which includes the bulk contribution $\hat H_\sigma$ and 
the hole kinetic term $\hat K$ which is written in terms of $T_f$. 
Collecting the various terms together our final expression for the
low energy spectrum of one hole at momentum $p$ is:
\begin{eqnarray}
E_p&=&E_\sigma+E_h \label{toten}\\
E_\sigma&=&e_0\ell-v_s{\pi\over 6\ell}+
{2\pi v_s\over \ell}\left({(n+\gamma)^2\over 4 K_\sigma}+K_\sigma
(m+\delta)^2\right ) \label{magen}\\
E_h&=& \epsilon_h (k_h) \label{holen}\\
k_h&=&p+Q\label{conserva2}\\
Q&=& k_F\left (2m+1\right)
+ {2\pi\over\ell} \left[ (\omega_++n+\gamma)(m+\delta)+ {\omega_- \over 2 }
 (n+\gamma)  \right]  \label{fsq} 
\label{moment}
\end{eqnarray}
where we have defined $\omega_\pm=(k_F^+\pm k_F^-){\ell\over 2\pi}$ (mod $1$).
In the thermodynamic limit, the quantization rule of 
the spinon momentum reduces to the non interacting result which, 
for odd $\ell$, reads
\begin{equation}
Q_m=k_F\left(2m+1\right )
\label{qm}
\end{equation}
but the finite size corrections of energy and momentum explicitly depend
on the phase shifts.
We stress that both the parameters entering the effective, long wavelength 
Hamiltonian ($v_s$, $K_\sigma$, $k_F^{\pm}$) and the phase shifts
($\delta_{\pm}$) cannot be trivially related 
to the bare lattice Hamiltonian because the RG flow 
renormalizes all the couplings not  protected by conservation laws. 
Their value can however be uniquely determined in Bethe Ansatz soluble
model by matching the form of the finite size corrections of the
one hole energy and momentum. This program will be pursued in the following 
Sections. More information can be gained at zero magnetization.
In this case, the effective spin model (\ref{hamilt}) has the
additional $SU(2)$ spin rotational symmetry which limits the
possible boundary conditions of the Luttinger model. In fact, in
spin isotropic models, the allowed boundary conditions are either
open or periodic: non trivial phase shifts are not compatible with the 
requirement of spin isotropy. Therefore, according to our basic assumption
the hole kinetic term stabilizes the periodic boundary conditions
in the effective spin Hamiltonian. This observation leads to a 
unique determination of the phase shifts $\delta_\pm$ 
at zero magnetization which only depend on the parity of 
the number of sites $\ell$. We first note that 
the $z$ component of the total spin $S^z$ is simply
related to the total number of fermions and then the spin excitations
are labeled by the quantum number $n$ in (\ref{nm}). By definition
$n=0$ corresponds to the ground state.
Moreover, in zero magnetic field, states with opposite values of 
$S^z$ are degenerate. For even $\ell$, the ground state is a singlet
and then excited states labeled by $n$ and $-n$ are degenerate: This
implies $\gamma=0$ through Eq. (\ref{magen}). For odd $\ell$ the ground state
itself is a spin doublet which gives $\gamma=1/2$ (mod 1) by the
same argument. For periodic boundary conditions also the total momentum
is a good quantum number. At fixed magnetization, 
excited states of definite momentum are
labeled by $m$ in Eq. (\ref{nm}), the ground state corresponding to the
choice $m=0$. For even $\ell$ the ground state is unique and
excited states with opposite momentum are degenerate giving $\delta=0$
by (\ref{magen}) while, for odd $\ell$ the ground state has finite momentum 
which implies a twofold degeneracy due to parity. Such a degeneracy is
compatible with the form of the energy spectrum (\ref{magen})
only for $\delta=1/2$. 

As an example of the previous analysis, we plot in Fig. 2 the overlap square
 $Z$ between the Heisenberg
ground state on a $\ell=L-1$ site ring and the ground state of a hole 
of momentum $p=\pi/2$
in the $L$ site $t-J$ model at $J=4t$ and vanishing magnetization.
The results have been obtained by Lanczos diagonalizations in 
chains with even $L \le 26$. According to our analysis we expect that at
long wavelengths the single hole problem in an {\sl even} chain is described by
an effective Heisenberg Hamiltonian with $\delta=\gamma=1/2$ which correspond
to {\sl periodic} boundary conditions on $L-1$ sites,
leading to a finite overlap in the thermodynamic limit.
Actually, the numerical results provide quite a strong evidence in
favor of this picture showing an overlap $Z$ which 
{\sl increases} with the size of the system.

In conclusion we see that spin isotropy
determines the quantization constant $\gamma$ while the occurrence of
periodic boundary conditions in the effective long wavelength Hamiltonian
fixes the value of $\delta$. At finite magnetization, spin isotropy
is broken and we expect the occurrence of generic, momentum dependent
values for $\gamma$ and $\delta$. These phase shifts will play an 
important role in determining the singularities of the Green function.
This subject will be discussed in the following Section.

For completeness let us briefly extend the previous discussion of 
the energy spectrum of the model to the case of {\sl open} boundary
conditions. According to our assumption this case is relevant only
when the hole effective hopping amplitude $t$ vanishes. 
The general form of the
bulk Hamiltonian $\hat H_\sigma$ (\ref{defh0}) does not depend on
the choice of boundary conditions and is therefore unaltered. However,
after having performed the canonical transformation (\ref{period}),
the normal modes are now defined by use of standing waves
rather than the previously introduced plane waves (\ref{planew}):
\begin{eqnarray} \label{stand}
\Psi_n&=& \sqrt{2 \over \ell} \int_0^l \sin (k_n x ) \Psi(x)
=\sqrt { K_\sigma \over 2 k_n } 
(a^{\dagger}_{k_n}+ a_{k_n}) \nonumber \\
\Pi_n &=&=  \sqrt{2 \over \ell} \int_0^l \sin (k_n x ) \Pi(x) 
=i\sqrt { k_n\over 2 K_\sigma  } (a^{\dagger}_{k_n}- a_{k_n})
\end{eqnarray}
where now $k_n={ (2 n + 1) \pi \over 2 \ell } $ with $n \ge 0.$
The bulk Hamiltonian for open boundary conditions then reads:
\begin{equation} \label{diagopen}
\hat H_\sigma= E_0 +  v_s \sum_{n\ge 0} k_n a^{\dagger}_{k_n} a_{k_n}
+{v_s\over 2\ell K_\sigma}(n+\gamma)^2
\end{equation}
The final expression for the
energy spectrum of the hole problem at $t=0$, i.e. when open boundary 
conditions apply, is then
\begin{equation}\label{fsopen}
E = E_0+ { 2 \pi v_s \over \ell } { 1 \over 4  K_\sigma}  (n+\gamma)^2
\end{equation}
where now the size scaling of the reference energy is \cite{affleck}
\begin{equation}
E_0 =  e_0 \ell - v_s {  \pi \over 24\, \ell } 
\label{centralopen}
\end{equation}
Again, in the spin isotropic case the values of the phase shifts
are constrained by the $SU(2)$ symmetry which gives $\gamma=0$
for even $\ell$ and $\gamma=1/2$ for odd $\ell$.

Now we conclude this Section by expressing the expected finite size corrections
to the one hole energy which emerge from our picture. On the basis of the
discussion at the beginning of this Section, it is particularly convenient
to work at fixed holon momentum $k_h$ extracting the size corrections of
the low energy excitation spectrum obtained by varying the spinon 
momentum and the magnetization of the model. These results will
be later compared with the exact form of the energy in two Bethe ansatz 
soluble models. The size dependence of the hole kinetic contribution 
(\ref{holen}) can be obtained by direct substitution of 
the explicit form of the spinon momentum (\ref{moment}) giving a term
proportional to the charge velocity
\begin{equation}
v_c(k_h) = {\partial \epsilon_h (k_h)  \over \partial  k_h }
\label{charge}
\end{equation}
Here $k_h$ is related to $Q_m$ and to the total momentum $p$ by the
conservation law (\ref{conserva2}). We also allow for a size dependence of 
the total hole momentum $p$: 
\begin{equation}
p=p_0+{2\pi\alpha\over \ell}
\label{momscal}
\end{equation}
valid up to $o(1/\ell)$ terms.
The constant $\alpha$ depends on the adopted sequence of lattice sizes.
By the conservation law $k_h=p+Q$ (Eq. \ref{conserva2}) the holon 
momentum acquires the finite size corrections of  
$p$ (Eq. \ref{momscal}) and $Q$ (Eq. \ref{fsq}). Therefore,
the holon kinetic term (\ref{holen}) will contribute to the $O(1/\ell)$ 
size corrections of the ground state energy as
\begin{equation} \label{masterc}
\Delta E_h = {\displaystyle 2 \pi v_c\over \ell} \left [ \alpha +
(\delta+m) (\gamma+n) + \omega_+ (\delta+m) +
{1\over 2}\omega_- (\gamma+n) \right ]
\end{equation}
On the other hand, the spinon term, being described by the conformal 
field theory which characterizes the Luttinger liquids, gives rises to the 
finite size corrections already obtained in Eq. (\ref{magen}).

In conclusion, at fixed holon momentum $k_h$, 
the finite size corrections of the energy $E_p$ of the
single hole problem give rise to a tower of states 
which depends on two quantum numbers $(n,m)$ and have
total momentum $p=k_h-Q_m$ (\ref{qm}). The size corrections are uniquely
determined by three bulk properties of the system 
($v_c$, $v_s$, $K_\sigma$) and four additional constants which
determine the boundary conditions of the effective spin Hamiltonian
($\omega_{\pm}$, $\gamma$, $\delta$):
\begin{eqnarray} \label{master}
\Delta E_p  = &{\displaystyle 2 \pi v_c\over \ell} &\left [ \alpha +
(\delta+m) (\gamma+n) + \omega_+ (\delta+m) +
{1\over 2}\omega_- (\gamma+n) \right ] \nonumber \\
+ &{\displaystyle  2 \pi v_s \over \ell}&
\left [ K_\sigma (\delta+m)^2 +{1 \over 4 K_\sigma} (\gamma+n)^2 \right ]
-{\pi\over 6\ell}v_s
\end{eqnarray}
Through this equation we can in principle evaluate the elusive 
spinon phase shifts $\delta_\pm$ by computing the easily accessible
finite size corrections of the one hole energy.
This equation has been obtained in the framework of the $t-J$ model
where no double occupancy is allowed but we will show that the same 
structure persists also in the one dimensional Hubbard model at
finite $U$. Therefore we believe that this form of the finite size
scaling is a general feature of one dimensional correlated models.

\subsection{Orthogonality catastrophe and the hole Green function}
\label{greensec}

In this Section we like to relate the previously introduced phase shifts
with the behavior of physical quantities and specific dynamical 
correlation functions of the one hole problem. In particular, we
will address first the evaluation of the hole quasiparticle weight 
at holon momentum $k_h$ and then the calculation of the asymptotic 
behavior of the Green function. 

The quasiparticle weight is defined as the square of the modulus of the 
matrix element:
\begin{equation}
\zeta=<\,k_h\,|\,c_{p,\downarrow}\,|\,\Psi\,>
\label{zetadef}
\end{equation}
where $|\,k_h\,>$ is the exact one hole ground state at holon momentum $k_h$,
$|\,\Psi\,>$ is the ground state of the Heisenberg model and we have chosen
the convention of creating a hole with $\downarrow$ spin projection. 
Momentum conservation implies that in the ground state $k_h=p\pm k_F$,
where $k_F$ is the spinon Fermi momentum.
By use of the Galileo transformation on the hole problem and a further
Jordan-Wigner transformation on the ``up" spins, 
the problem is reduced to the evaluation of the overlap
between the two fermionic states corresponding to the Heisenberg
ground state $|\,\Psi\,>$ on a $L$ site ring and the ground state $|\,k_h\,>$
of the effective spin Hamiltonian defined on
the lattice of $\ell=L-1$ sites 
and the same number of up spins. 
In order to study the behavior of the quasiparticle weight for
$L\to \infty$, we can limit our attention to the long wavelength form
of the effective spin Hamiltonian.
The problem is therefore to compute the overlap of two eigenstates
of the same bosonic Hamiltonian (\ref{defh0}), with
different boundary conditions: while the reference state $|\,\Psi\,>$ has
standard periodic boundary conditions (i.e. $\Pi^*=\Delta\Phi=0$), 
the one hole ground state $|\,k_h\,>$ is defined by non universal, 
momentum dependent, values of the phase shifts:
\begin{eqnarray}
\Pi^* &=& 2 \sqrt{\pi}\, \delta 
=\sqrt{\pi}\,(\delta_++\delta_-)\nonumber \\
\Delta \Phi&=& \sqrt{\pi} \,\gamma=\sqrt{\pi}\,(\delta_+-\delta_-)
\label{peff}
\end{eqnarray}
In the continuum limit this overlap is strictly zero because the two states 
are eigenstates of the bosonic Hamiltonian with different eigenvalues for
$\Delta \Phi $ and $\Pi^*$.
This orthogonality is however an artifact of the Luttinger
extension to a system containing an infinite number of particles and
it is clear that in a finite system the
overlap between two states with different boundary conditions will be
in general finite.
The solution is to ``regularize'' the  Luttinger branches, in order
to be consistent with a tight binding model where the same phase shifts
$\delta_+$ and $\delta_-$ at the Fermi energy are induced by a  local
potential which gives no contribution to the backward scattering.
In a free Fermi gas, $K_\sigma=1$ and the two states
can be written as Slater determinants of plane waves with suitable phase 
shifts. In this case, the overlap can be easily calculated giving an
asymptotic power law behavior which only depends on the value 
of the phase shifts at the Fermi energy of the right and left branches:
\begin{eqnarray} 
\zeta&=&\,
<\, k_h\, |\, \Psi\,> \, \propto \exp\left [\displaystyle -{1 \over 2 }
\left ( \delta_+^2 +\delta_-^2\right ) \ln \ell \right ]\,\sim L^{-X_0}
\label{zpower} \\
X_0&=&{1\over 2} \left (\delta_+^2+\delta_-^2\right ) =   
(\delta^2 + { \gamma^2  \over 4}  )
\label{xexpo}
\end{eqnarray}
where the last equality in Eq. (\ref{xexpo}) follows from Eq.(\ref{peff}).
The formal calculation of the overlap is contained in appendix \ref{overapp}.
This formula generalizes the exact result valid for $\delta_+=\delta_-$ 
\cite{parola,penc} and agrees with the prediction of conformal field theory
relating the finite size corrections of the energy
to the exponent of the boundary operators \cite{affleck}. 
In fact, the explicit expression (\ref{xexpo}) coincides with
the term proportional to $2\pi v_s/\ell$ in the final formula
Eq. (\ref{master}) for the finite size corrections of the one hole energy.

This formalism can be easily generalized to the case $K_\sigma\ne 1$:
In fact, the overlap does not change upon unitary transformations
and it is known that the scaling of the bosonic fields
\begin{eqnarray} \label{transf}
\Pi^\prime (x)&=& \sqrt{K_\sigma}\, \Pi (x)  \nonumber \\
\Phi^\prime(x)&=& {1 \over \sqrt{K_\sigma}}\, \Phi (x)
\end{eqnarray}
maps the interacting problem with $K_\sigma\ne 1$ to the one with
$K_\sigma=1$. By the same transformation,
the boundary condition for the one hole state are modified due to 
Eqs. (\ref{quantumno},\ref{nm}): $\gamma^\prime=\gamma/\sqrt{K_\sigma}$,
$\delta^\prime=\delta\sqrt{K_\sigma}$,
while for the reference state the boundary condition are
unchanged. The effective phase shifts of the parent non interacting 
case then follow immediately:
\begin{eqnarray}\label{fondam}
\delta_+^{\rm eff} &=& {1\over 2}\left [
2 \sqrt{K_\sigma}\, \delta + { 1\over \sqrt{K_\sigma}}\,\gamma
\right ]  \nonumber \\
\delta_-^{\rm eff}&=& {1\over 2} \left [2 \sqrt{K_\sigma} \,\delta - 
{ 1\over \sqrt{K_\sigma}} \,\gamma \right ]
\end{eqnarray}
leading to the general expression:
\begin{equation}
X_0 =  \left[ K_\sigma \delta^2 + {1\over 4 K_\sigma } \gamma^2 \right]
\label{xs}
\end{equation}
again consistent with the finite size corrections to the energy (\ref{master}).

The relation between the
finite size corrections to the energy and the exponent $X_0$
of the orthogonality catastrophe is a general property of all conformal
field theories \cite{affleck} and holds also in other cases.
For instance, the exponent associated to the open boundary fixed point 
can be determined in terms of the finite size corrections obtained in
Eq. (\ref{fsopen}): 
\begin{equation}
X_0= { \gamma^2 \over 4 K_\sigma} + { 1\over 16}
\end{equation}
where use has been made of
the known additional contributions to the ground state energy with
periodic (\ref{centralpbc}) and open (\ref{centralopen}) 
boundary conditions on a $L$ and $\ell$ site chain respectively
\cite{affleck}. This expression, which applies 
in the limit of a static hole,
yields the exact exponent $X_0=3/16$ in the isotropic case, when
$K_\sigma=1/2$ and $\gamma=1/2$. This has been checked numerically in Fig. 3
by Lanczos diagonalization of the hole problem in the
limit of vanishing hopping amplitude $t$. 

Now we can proceed to the study of the asymptotic behavior of the 
retarded single hole Green function:
\begin{equation}
G_\sigma(p,\tau)=i\,<\,\Psi\,|c^\dagger_{p,\sigma}e^{-i(\hat H-i\eta)\tau} 
c_{p,\sigma}\,|\,\Psi\,>
\label{greendef}
\end{equation}
valid for $\tau >0$ with $\eta=0^+$ as convergence factor.
As usual, by performing the Galileo transformation and then a Jordan-Wigner
transformation we can map the problem of the evaluation of the Green 
function to the calculation of the purely fermionic matrix element:
\begin{eqnarray}
G_\uparrow(p,\tau)&=&i\,<\,\Psi\,|e^{-i(\hat H_p-i\eta)\tau} n_0\,|\,\Psi\,> 
\label{gup} \\
G_\downarrow(p,\tau)&=&i\,<\,\Psi\,|e^{-i(\hat H_p-i\eta)\tau} 
(1-n_0)\,|\,\Psi\,> 
\label{gdown}
\end{eqnarray}
where the effective Hamiltonian $\hat H_p$ is defined in Eq. (\ref{htot})
and the fermion projection operators $n_0=\psi_0^\dagger\psi_0\,$, 
$(1-n_0)=\psi_0\psi_0^\dagger$ force the spin at the origin of the chain to 
point upwards or downwards respectively. In the following, we will first
carry out the calculation of the asymptotic behavior of the
trace of the Green function matrix:
\begin{equation}
G(p,\tau)=G_\downarrow(p,\tau)+G_\uparrow(p,\tau)=
i\,<\,\Psi\,|e^{-i(\hat H_p-i\eta)\tau}\,|\,\Psi\,> 
\label{gtrace}
\end{equation}
which is expected to show all the singularities present in the two
separate spin projections. 

The  long wavelength form of the Hamiltonian $\hat H_p$ has been already studied
in the previous Section. $\hat H_p$ 
depends on the total momentum $p$ via a function
$\epsilon_h(k_h)$  which represents  the holon dispersion:
\begin{equation} \label{fixedham}
\hat H_p = \epsilon_h(p+\hat P)  + \hat H_\sigma
\end{equation}
$\hat H_\sigma$ is a Luttinger liquid Hamiltonian
characterized by suitable boundary conditions and
commuting with the spinon momentum operator $\hat P$ 
defined on the chain of $L-1$ sites. Note that the Heisenberg state 
$|\,\Psi\,>$ instead refers to the $L$-site ring with periodic boundary 
conditions and therefore
is not an exact eigenfunction of the spinon momentum operator $\hat P$.
It is then convenient to express the state 
$|\,\Psi\,>$ as the sum of all its projections into subspaces
of definite momentum $\hat P=Q$ and substitute this representation into
the form (\ref{gtrace}):
\begin{eqnarray} 
G(p,\tau)&=&  {i\over \ell}\,\sum_Q Z(Q,\tau) 
e^{ -i [\epsilon_h( p+Q)-i\eta]\,\tau } \label{convola} \\
Z(Q,\tau)&=& \int_0^\ell e^{-i Q x } Z(x,\tau)  dx \nonumber  \\
Z(x,\tau) &=& <\,\Psi\,|\,  e^{i  ( \hat P x -\hat H_\sigma \tau) }\,
|\,\Psi\,>
\label{defz}
\end{eqnarray}
This is the general form of the one hole Green function which shows the
effects of spin-charge decoupling on the dynamics of the hole.
The spinon function $Z(Q,\tau)$ provides a generalization of the quantity $Z(Q)$
which characterizes the form of the single hole 
Green function in the $J\to 0$ limit \cite{parola,penc}.
Here, however, non trivial dynamics of the spins are induced by $\hat H_\sigma$
as long as $v_s\ne 0$. 

Let us first discuss the calculation of the matrix element $Z(x,\tau)$
in the non interacting limit. In a free Fermi gas with arbitrary 
boundary conditions, the long wavelength Hamiltonian reduces to the sum 
of the kinetic terms for the two branches of right and left moving fermions.
In the continuum limit the Hamiltonian and the momentum operator
are given by:
\begin{eqnarray}
H &=& v_s(\hat P_+ - \hat P_-)\,+\,{\rm const} \nonumber\\
\hat P &=& \hat P_+ + \hat P_- \nonumber\\
\hat P_\pm &=& \pm k_F\int_0^\ell dx\psi^\dagger_\pm(x)\psi_\pm(x)+
i\,\int_0^\ell dx \,\psi^\dagger_\pm(x)\partial_x\psi_\pm(x) 
\label{hfree}
\end{eqnarray}
This particular form of the Hamiltonian shows a close relationship 
between energy and momentum operators which is clearly valid {\sl only} 
in the non interacting limit. However, the more general interacting 
Luttinger liquid can be mapped to the free Fermi gas by the 
previously defined canonical transformation (\ref{transf}) 
leading to the conservation
of the number of right and left moving fermions at long wavelength. 
As a consequence, all eigenstates would factorize in the product of two 
states, one for each branch and, via Eq. (\ref{defz}) 
also the function $Z(x,\tau)$ would split in the product of two 
terms $Z_\pm^{\delta_\pm}(x\mp v_s\tau)$ defined in each branch.
However, in the microscopic model, defined on a lattice,
higher order terms allow the excitation of a fermion from one branch 
to another (even for $K_\sigma=1$). Therefore, the function $Z(x,\tau)$ will 
contain contributions corresponding to all these excitations which 
we label by the number $m$ ($-m$) of extra fermions on the right 
(left) branch:
\begin{eqnarray} 
Z(x,\tau) &=& \sum_{m=-\infty}^{\infty}\,e^{ i (Q_mx - E_m\tau ) } \,
Z_+^{(\delta_+,m)}(x-v_s\tau)\,Z_-^{(\delta_-,-m)}(x+v_s\tau)
\nonumber\\
&=& \sum_{m=-\infty}^{\infty}\,e^{ i (Q_mx - E_m\tau ) } \,
<\,e^{i(\hat P_+-Q_+(m))\,(x -v_s\tau)}\,>_+\,
<\,  e^{i(\hat P_- -Q_-(-m))\,(x +v_s\tau)}\,>_-
\label{prodz}
\end{eqnarray}
where the average labeled by $+$ ($-$) is taken on the ground state 
of the right (left) branch in the undoped system and the 
intermediate states are constrained to have $m$ ($-m$) additional
particles on the right (left) branch. $Q_m=Q_+(m)+Q_-(-m)$ and
$E_m=v_s(Q_+(m)-Q_-(-m))$ represent the reference momentum 
and energy of the intermediate states with $m$-particle excitations 
which are explicitly given in terms of the Fermi momentum of
the spinons by $Q_+(m)=k_F(m+1)$ and $Q_-(-m)=-k_F(-m)$.
The functions $Z_{\pm}^{(\delta_{\pm},\pm m)}(x,\tau)$ introduced in Eq. 
(\ref{prodz})
are well defined also in the long wavelength limit where the spinon 
Hamiltonian $\hat H_\sigma$ can be written in the bosonized form 
(\ref{defh0}). As discussed in Sections \ref{bosesec} and \ref{boundsec}
the quantum number $m$ which characterizes the inter-branch excitations
appears in the low energy Hamiltonian only through the quantity 
$\Pi^*$ and can be absorbed in the definition of the phase shifts:
$\bar\delta_+=\delta_++m$, $\bar\delta_-=\delta_-+m$ 
(see Eqs. \ref{nm}, \ref{gam}, 
\ref{delt}). The calculation of $Z(x,t)$ then reduces to the 
evaluation of the contribution appropriate for each branch with
arbitrary phase shift $\bar\delta_\pm$. The technical details are discussed in
appendix \ref{greenapp}, here we just report the asymptotic behavior 
in the thermodynamic limit (\ref{finalz}):
\begin{equation}
Z_\pm^{\bar\delta}(x) \propto  (\mp i x )^{-\bar\delta^2 } 
\label{asintoz}
\end{equation} 
By substituting this result into Eq. (\ref{defz}),
the spectral weight $A(p,\omega)= { 1 \over \pi } {\rm Im } G(p,\omega)$
is written in terms of the ($Q,\omega$) Fourier transform of $Z$:
\begin{eqnarray} 
A(p,\omega) &=& \int { dQ \over 2 \pi }
Z(Q, \omega -\epsilon_h(p+Q) ) 
\label{spectralw} \\
Z(Q,\omega)&=&\int_{-\infty}^{\infty} \int_{-\infty}^{\infty} \,d\tau d\,x
\,e^{ - i ( x Q - \omega \tau ) } Z(x,\tau)   \nonumber \\
&=& \sum_m C_m \, \Theta \left [ \omega + v_s (Q-Q_m) \right ]  \,
\Theta \left [ \omega -v_s (Q-Q_m) \right ]
\nonumber \\
& &v_s^{1-2X_m}\,
\left [\omega+v_s(Q-Q_m) \right ]^{\bar\delta_+^2 -1}
\left [\omega-v_s(Q-Q_m) \right ]^{\bar\delta_-^2-1}
\label{zqom}
\end{eqnarray}
In the Fourier transform of $Z(x,\tau)$ only the singular
contributions have been included. 
$\Theta(x)$ represents the step function whose presence is a 
direct consequence of the Fourier positivity (negativity) of the 
function $Z_+$ ($Z_-$), as discussed in appendix \ref{greenapp}. 
This property  has a simple physical meaning:
The particle hole excitations within  the right branch of a 
Luttinger model can only increase the total momentum with respect to the 
ground state, while the excitations on the left branch 
can only decrease the total momentum of the system. Therefore the
spinon spectral function at $Q>0$ has contributions coming only from 
excitations in the right branch and {\it vice versa}.
The constants $C_m$, which only depend on the phase shifts $\delta_+$
and $\delta_{-}$  in a symmetric fashion,
can be explicitly calculated in the free Fermi gas 
with skew boundary conditions as shown in Eq. (\ref{zqgen}) of
appendix \ref{greenapp}. However, we expect that in the interacting models 
these coefficients will be renormalized in a non universal way which 
depends on the physical cut-off present in the microscopic model.

By inserting Eq. (\ref{zqom}) into Eq. (\ref{spectralw}), the 
contribution coming from each excited state $m$ gives a divergence
in the spectral function as long as $\bar\delta_+^2+\bar\delta_-^2 < 1$.
The singularities are located along the lines 
\begin{equation}
\Delta \omega_m =\omega - \epsilon_h(p+Q_m)=0 
\label{locus}
\end{equation}
in the ($p,\omega$) plane and show the asymptotic behavior:
\begin{equation}
A(p,\omega) \propto  |\Delta \omega_m |^{ 2 X_m -1} 
\label{singa}
\end{equation}
where $X_m={1\over 2}(\bar\delta_+^2+\bar\delta_-^2)$. 
In general, we expect that the phase shifts
$\delta_\pm$ which characterize the singularities of the spectral
function depend on the holon momentum $k_h$. Therefore, 
$A(p,\omega)$ will show singularities along the lines 
$\omega = \epsilon_h(p+Q_m)$ with momentum dependent exponents $X_m$.

Another unexpected prediction of the present formalism concerns
the behavior of the spectral function in a neighborhood of the singularity:
By a direct evaluation of the prefactor we find 
different results according to whether the charge velocity is larger or 
smaller than the spin velocity. In fact,
if $|v_c| \le v_s$, there are divergences only for 
$\omega \ge \epsilon_h (p+Q_m)$ while the spectral weight vanishes on 
the other side of the singularity line $\Delta\omega_m=0$ as there are no 
states with lower energy contributing to the spectral weight:
\begin{equation}
A(p,\omega ) \propto \,{|\Delta \omega_m|^{2 X_m  -1} 
\over (v_s +v_c)^{\bar\delta_+^2} (v_s-v_c)^{\bar\delta_-^2} }
\label{singa1}
\end{equation}
Instead if $|v_c| > v_s$ the spectral function diverges on both sides of 
the singularity with the same exponent $2 X_m -1$ but different prefactor:
\begin{equation}
A(p,\omega)\propto\, { \pi \over \sin(\pi \bar\delta_\mp^2)  }\,
{|\Delta \omega_m|^{ 2 X_m -1} \over |v_s+v_c|^{\bar\delta_+^2} 
|v_s-v_c|^{\bar\delta_-^2 }  } 
\label{singa2}
\end{equation}
where the upper sign and the lower one refer to $\pm v_c\Delta\omega >0$, 
 respectively.
The amplitude ratio can be evaluated from Eq. (\ref{singa2}):
\begin{equation} \label{univratio} 
{A(\Delta\omega)\over A(-\Delta\omega)}=
{\sin(\pi\delta_{\pm}^2)\over\sin(\pi \delta_\mp^2)}
\end{equation}
This simple expression, which 
shows the asymmetry of the spectral weight above and below the singularity, 
is strictly valid only for the Luttinger model, but is
also expected to be qualitatively correct in more general cases.
In fact this feature originates from the different energy 
spectrum of the excitations on the left and right branches, which is
a common property of one dimensional correlated systems.
Notice also that the prefactors in (\ref{singa1}, \ref{singa2}) 
are strongly enhanced close to the instability 
which sets in when the charge velocity equals the spin velocity.

When we switch on interactions among spinons the above expressions
(\ref{spectralw}, \ref{zqom}, \ref{singa1}, \ref{singa2}) 
remain formally unchanged if the phase shifts are suitably 
renormalized by $K_\sigma$ analogously to Eq. (\ref{fondam}) where 
$\delta_\pm^{\rm eff}\to\bar\delta_\pm$ and $\delta\to(\delta+m)$.
As a consequence, the exponent of the leading singularity (i.e. 
that corresponding to $m=0$) $X_0$ exactly reproduces the exponent
already obtained in the calculation of the overlap $\zeta$ (\ref{xs}).
Moreover, the full set of exponents $X_m$ agrees with the term 
proportional to $2\pi v_s/\ell$
in the finite size correction to the energy of 
the model (\ref{master}) for the particular choice of the quantum number $n=0$.
The restriction $n=0$ is due to the fact that the intermediate
states appearing in the rotational invariant Green function 
(\ref{gtrace}) do not have definite spin and then, at fixed $m$, the leading 
singularity is related to the smallest phase shift which corresponds to 
$n=0$. It is however
clear how to generalize these expressions for the calculation of 
the spin up (or spin down) Green function: The sum over intermediate states 
in Eq. (\ref{convola}) has to be restricted to the states with the correct
spin projection. Within bosonization it means states with the appropriate 
value of the quantum number $n$ 
which is in fact related to the total spin of the one hole 
intermediate state through Eqs. (\ref{quantumno},\ref{nm}). This 
procedure in fact reproduces the full set of critical exponents $X_{m,n}$
which appear in Eq. (\ref{master}).

Finally, if we want to evaluate the singularities in the hole density of
states we just have to integrate the leading momentum dependence of the
spectral function (\ref{singa}):
\begin{equation}
N(\omega)=\int {dp\over 2\pi} A(p,\omega) \propto 
\sum_m c_m \,\omega^{2X_m(0)-1/2} + {\rm const} 
\label{dos}
\end{equation}
where $c_m$ are finite amplitudes, $\omega$ is measured from the 
bottom of the band and
the exponent $X_m(0)$ coincides with the previously introduced 
critical exponent $X_m$ evaluated {\it at the bottom of the holon band}
i.e. at a total momentum $p$ such that $k_h=p+Q_m$ sits at the
minimum of $\epsilon_h(k_h)$. In fact this region of integration 
in momentum space gives rise to the leading singularity in the density of 
states. Note that a divergence in the density of 
states occurs only if $X_m(0) < 1/4$ for some $m$.

\setcounter{equation}{0}

\section{Single hole in the XY model}
\label{txysec}
Here we analyze in some detail the dynamical properties of a single hole in the
$XY$ model on the basis of the Hamiltonian already introduced in 
Section \ref{impurasec} in the limit of small hole mass: $J^\prime=J << t$.
The aim of this study is to check, in a simple model, all the general 
features of hole motion already discussed in Section \ref{fieldsec}
and to carry out the quantitative evaluation of exponents and amplitudes 
for this system. 

The Hamiltonian of a hole of momentum $p$ in the $t-J_{XY}$ model 
(\ref{wigj}) has been previously derived in some detail and reads: 
\begin{equation}
\hat H_p =t\,\exp \left [ip+i\sum\limits_{k}
k\, \psi^{\dagger}_k \psi_k\right ] + h.c.
-{J\over 2} \sum_{i=1}^{\ell-1} 
\left [\psi^{\dagger}_i \psi_{i+1} + h.c.\right ]
\label{hspinon}
\end{equation}
Here we consider a {\sl spin down} hole of momentum $p$ in a
chain with even number of sites $L$ and odd number of up spins (i.e. 
an odd number of fermions in the representation of Eq. 
(\ref{hspinon})) $N=2\nu+1$ corresponding to a $z$-axis magnetization 
$\mu=-{1\over 2}+{N\over L}$. According to the discussion in Section 
\ref{effsec}, the appropriate boundary conditions 
of the fermionic problem are therefore {\sl antiperiodic}, $N+L$ being odd,
and the quantization rule for the momenta is 
\begin{equation}
k_j={2\pi\over \ell}(j+{1\over 2})
\label{kj}
\end{equation}
with $\ell=L-1$ as usual. The states which diagonalize the hole kinetic
term are states of given spinon momentum $Q$. In the $J\to 0$ limit
the spectrum of the Hamiltonian (\ref{hspinon}) can be obtained by
diagonalizing the magnetic term in the subspace of fixed spinon
momentum $Q$. This procedure gives rise to spinon Slater determinants
of plane waves. The ground state is doubly degenerate and
corresponds to a set of $N$ occupied orbitals of 
momenta centered around $k=0$, while low energy excitations
can be obtained either 
by changing the number of fermions (i.e. the magnetization)
$N\to N+n$ or by moving $m$ fermions from the left to the right branch
of the Fermi surface. The energy of these excited states
can be easily calculated by taking the expectation value of the 
magnetic term on the appropriate spinon Slater determinant. Energy 
and momentum are therefore given by:
\begin{eqnarray}
E_p&=&2t\cos(p+Q) - J{\ell-1\over \ell}\sum_{j=-\nu-n+m}^{\nu+n+m} 
\cos\left[{2\pi\over \ell}(j+{1\over 2})\right]
\label{ener}\\
Q&=&\sum_{j=-\nu-n+m}^{\nu+n+m} {2\pi\over \ell}(j+{1\over 2})
\label{mom}
\end{eqnarray}
These expressions, being based on perturbation theory in $J << t$, are exact
to $O(J/t)$ and to 
$O(1)$ respectively. By carrying out the summations and expanding
up to $O(1/\ell )$ at fixed $\rho=N/L$, we find:
\begin{eqnarray}
E_p&=&-J(\ell-1) {\sin\pi\rho\over\pi} -J\cos\pi\rho\, (n+\rho) 
+2t\cos(p+Q) +\nonumber \\
&\phantom{=}& 
{2\pi J\sin\pi\rho \over \ell} 
\left [(m+{1\over 2})^2 + {1\over 4}(n+\rho)^2\right ] 
-J\sin\pi\rho{\pi\over 6\ell} +
{J\cos\pi\rho \over \ell}(n+\rho) 
\label{ener2}\\
Q&=&2\pi\rho(m+{\textstyle{1\over 2}}) +{2\pi\over\ell}(n+\rho)(m+
{\textstyle{1\over 2}})
\label{mom2} 
\end{eqnarray}
In the thermodynamic limit, besides the extensive magnetic contribution, 
the energy of the state depends on the holon momentum $k_h=p+Q$ through
the form of the holon band $\epsilon_h(k)=2t\cos k_h$. The spinon momentum
is instead given by $Q=2 k_F(m+{1\over 2})$ when we recognize that the spinon
Fermi momentum is just $k_F=\pi\rho$. The $O(1/\ell)$ size corrections to 
momentum and energy can be also compared to the general expressions 
(\ref{moment},\ref{master}) if we recall that the spinon velocity is 
$v_s=J\sin k_F=J\sin\pi\rho$ and $K_\sigma=1$ in the $XY$ model. 
Equations (\ref{ener2}) and (\ref{mom2})
correctly reproduce the predicted structure of the size corrections
showing that the hole dynamics is described by an underlying  conformal 
field theory. This comparison allows to determine, to lowest order in $J/t$,
the values appropriate for the $t-J_{XY}$ model
of the non universal parameters entering Eq. (\ref{master}):
\begin{equation}
\delta={1\over 2}(\delta_++\delta_-)={1\over 2}
\qquad\qquad\qquad \gamma=(\delta_+-\delta_-)=\rho 
\label{phashiftxy}
\end{equation}
for the phase shifts and 
\begin{equation}
\omega_+={\ell\over 2\pi} (k_F^++k_F^-)=0 
\qquad \omega_-= {\ell\over 2\pi} (k_F^+-k_F^-)= O(J)
\label{paramxy}
\end{equation}
for the shifts of the right and left Fermi wavevectors. 

According to Section \ref{greensec}
the phase shifts (\ref{phashiftxy}) completely determine the singularities
of the Green function. As a check, let us explicitly evaluate
the overlap $\zeta$ (\ref{zetadef}) in this  model. In fermionic
representation this amounts to calculate the overlap between 
the ground state of Hamiltonian (\ref{hspinon}) and
the ground state of the $XY$ model on a $L$-site ring 
with the same number of fermions $N=2\nu+1$. In fermion representation 
the latter state is just a Slater determinant of plane waves with momentum
quantization corresponding to {\sl periodic} boundary conditions on
a $L$-site ring: 
\begin{equation}
q_j={2\pi\over L}j 
\label{qj}
\end{equation}
Therefore, $\zeta$ is
simply the overlap of two Slater determinants of plane waves with different 
quantization rule (\ref{kj},\ref{qj}). Such an overlap can be calculated
as the determinant of the $N\times N$ matrix of the overlaps of the two sets of 
plane waves defined on the squeezed chain:
\begin{eqnarray}
\zeta_{p,\downarrow}&=& {\det H} \label{detz}\\
H_{rs}=<q_r\,|\,k_s>&=&i{e^{i k_s/2}\over\sqrt{L(L-1)}} 
{\cos\left [q_r /2\right ]\over\sin\left [(k_s-q_r)/ 2\right ]}
\label{mij}
\end{eqnarray}
with $r$ and $s$ belonging to the interval $[-\nu,\nu]$. Notice that 
the quantization rule of the wavevectors $k_j$ can be naturally interpreted
in terms of a momentum dependent phase shift:
\begin{equation}
k_j\sim {2\pi\over L}(j+{1\over 2}+{j\over L})
\end{equation}
Near the Fermi points the phase shifts are then given by: 
$\delta_\pm={1\pm\rho\over 2}$ in agreement with their determination based 
on the structure of the finite size corrections to the energy 
(\ref{phashiftxy}). The known treatment \cite{orto} of the 
orthogonality catastrophe problem thus gives the critical exponent
in terms of the phase shift at the Fermi points: $2X_0=
\delta_+^2+\delta_-^2$ which agrees with the expression (\ref{xexpo})
derived in Section \ref{greensec}. This result can be also 
checked numerically by evaluating the determinant (\ref{detz})
for fairly large system sizes.
The size scaling of $\ln{\zeta}$ as a function of $\ln N$ 
is shown in Fig.4 for two magnetizations ($\mu=0$ and $\mu=1/4$)
corresponding to the densities of spin up $\rho=1/2$ and $\rho=3/4$.
The analytical value of the exponent $X_0$ is also shown in figure.
The exponent for the $\rho=1/2$ case also agrees with an
independent calculation by Penc {\it et al.} \cite{penc2}.

The next task is the evaluation of the spin down Green function 
(\ref{greendef}) which, via Galileo transformation, takes the
form (\ref{gdown}). In the
$J\to 0$  limit, the energy levels can be written as the sum of a holon
part $\epsilon_h(p+Q)=2t\cos(p+Q)$,
which just depends on the spinon momentum $Q$,
and a spinon term which is eigenstate of the
$XY$ Hamiltonian on the squeezed chain. Therefore, at fixed total 
momentum $p$, the Green function reads
\begin{equation}
G(p,\tau)={i\over \ell}\sum_Q\sum_R e^{-iQR} \,e^{-i\epsilon_h(p+Q)\tau}\, 
Z(R,\tau)
\label{green3}
\end{equation}
where the spinon term, implicitly depending on the total momentum $p$, is
\begin{equation}
Z(R,\tau)=<\Psi\,|\, e^{-i\hat H_\sigma \tau} T_{\ell}^{R} (1-n_0)\,|\,\Psi>
\label{effe}
\end{equation}
Here  $\hat H_\sigma$ and $T_{\ell}$ respectively 
represent the $XY$ spinon Hamiltonian
and the translation operator of one lattice spacing 
on the squeezed ring of $\ell=L-1$ sites (origin excluded) while 
$|\Psi>$ is the ground state of the $XY$ model on the $L$-site ring. 
Notice that, for the special case of the $XY$ model at $J\to 0$, 
the decoupling of the Green function in holon and spinon factors
\begin{equation}
G(R,\tau)=Z(-R,\tau)\,G_h(R,\tau)
= Z(-R,\tau)\,i\,\int_{-\pi}^\pi {dk\over 2\pi}e^{ikR} 
\,e^{-i\epsilon_h(k)\tau} 
\end{equation}
is exact at all distances while, in general, we expect 
this decoupling  is valid only at low energy and long wavelength, 
i.e. for $R,\tau \gg 1$.

The evaluation of the matrix element in
Eq. (\ref{effe}) can be performed because in the spinon representation
the two sates are Slater determinants (without phase shifts) and the
unitary operator acting on them is a one body operator. The projection
operator $(1-n_0)$ implies that the origin is
an empty site which amounts to exclude the origin in the evaluation of
the overlap matrix. Therefore, $Z(R,\tau)=\det A(R,\tau)$ 
where the overlap matrix
(of linear dimension $2\nu+1$) can be easily calculated by inserting a
complete set of orbitals which do not place particles in the origin.
A useful choice is the set of eigenstates of the spinon Hamiltonian
$\hat H_\sigma$ corresponding to an odd number of spinons
i.e. with momenta (\ref{kj}). The resulting form of the overlap matrix is:
\begin{equation}
A_{rs}(R,\tau)={\cos(q_r/2)\cos(q_s/2)\over L(L-1)}\sum_{j=-L/2+1}^{L/2-1}
{e^{i(k_jR +\tau J\cos k_j)}\over \sin[(k_j-q_r)/2]\,\sin[(k_j-q_s)/2]}
\label{overmat}
\end{equation}
The sum can be performed analytically only in the static limit $\tau=0$ 
where the numerical computation 
of $Z(R,0)$ and of its Fourier transform $Z(Q)$ can be
pushed to fairly large system sizes. A comparison between two different
sizes at fixed magnetization ($\mu=0$ and $\mu=1/4$) is shown in Fig. 5a
where we can see that $Z(Q)$ does not vanish outside the
``spinon Fermi surface" even if it is strongly suppressed. The
singularity at the Fermi momentum $k_F=\pi\rho$ appears to be present on
both sides of the spinon Fermi surface.
A logarithmic plot of the singularity of $Z(Q)$ when $Q$ approaches $k_F$
is shown in Fig. 5b where it is compared to the analytical value 
of Eqs. (\ref{phashiftxy}, \ref{zqgen}).
The ratio of the two amplitudes on both sides of the singularity is in rather 
good quantitative agreement with expression (\ref{univratio}). 

Now we examine in some detail the asymptotic form of the {\sl dynamical}
spinon Green function of the $t-J_{XY}$ model (\ref{effe}). As noted before,
the exact calculation of $Z(R,\tau)$ in a finite system reduces to the
evaluation of the determinant of the matrix $A_{rs}(R,\tau)$ defined in Eq.
(\ref{overmat}). The exact calculation can be performed only numerically
and the direct interpretation of the data is obscured by finite size
effects. However, we can address the problem of the long distance
and long time behavior
of $Z(R,\tau)$ by performing the asymptotic expansion of the
matrix elements $A_{rs}$ themselves. This expansion can be carried
out rather easily in the low density limit where, according to expectations
of conformal field theory, it should be characterized by 
phase shifts $\delta_+=\delta_-=1/2$ (\ref{phashiftxy}). The details are 
reported in appendix \ref{aptxy}.

According to bosonization, the spinon Green function should behave as
(\ref{asintoz})
\begin{equation}
Z(R,\tau)\propto
{e^{ik_F R}\over (R-v_s \tau)^{1/4} (R+v_s \tau)^{1/4}} +
{e^{-ik_F R}\over (R-v_s \tau)^{1/4} (R+v_s \tau)^{1/4}}
\label{boson}
\end{equation}
where $v_s=J\,k_F$ and this expression is valid for $|v_s\, \tau|<R$.
From this analysis we expect that the function $Z(R,\tau)$ at long
wavelength behaves as
\begin{equation}
Z_0(R,\tau)\propto
{\cos(k_F \, R)\over (R-v_s \tau)^{1/4} (R+v_s \tau)^{1/4}}
\label{detb}
\end{equation}
In Fig. 6 we plot the numerically evaluated ratio
$|\,\det A(R,\tau)\,| /Z_0(R\tau)$
as a function of $R-v_s\tau$ for $N=100$ and $\mu=-1/4$
which belongs to the low density regime.
This ratio has no oscillations (meaning that the phase factor
has been correctly determined) and it is approximately constant over a
wide range of values of $R-v_s \tau$. Clearly the region $R\sim v_s \tau$
cannot be well represented by the bosonized form which would
predict a spurious divergence in the Green function which instead is
bound to have modulus less than unity.

The analysis of this Section shows in a simple example that 
all the features of hole propagation in a magnetic background derived 
with field theoretical formalism are contained in such a microscopic 
model. This detailed calculation supports the assumptions introduced in our
general study of the long wavelength hole dynamics. More interesting
systems can now be investigated.

\setcounter{equation}{0}

\section{Bethe Ansatz models}
\label{hubtjsec}

In this Section we consider two Bethe ansatz solvable models,
the repulsive Hubbard model \cite{liebwu} and the $t-J$ model at $J=2t$
\cite{bares}, where the finite size corrections to the energy can be found 
analytically, leading to a formal expression for the critical exponent
which appears in the one hole Green function. We analyze both
models at arbitrary magnetization so that our results can be 
extended to the attractive Hubbard model via the well known canonical
transformation.

The calculation of the finite size corrections to the ground state
energy closely follows the original derivation by Woynarovich
for the Hubbard model at finite density \cite{woy}. 
Here we consider a chain of 
$L$ sites and $L-1$ electrons, among which $N_{\downarrow}$ have spin down.
The Bethe ansatz solution is characterized by two sets of rapidities: 
For the Hubbard case we have $N_s=N_{\downarrow}$ rapidities $\lambda_{\alpha}$ 
and $N_c=L-1$ rapidities $k_j$, while for the $t-J$ model (in Sutherland 
representation) we have $N_s=N_{\downarrow}+1$ rapidities $v_{\alpha}$ and
$N_c=1$ rapidity $w_0$. These rapidities are related to the quantum numbers
$J_{\alpha}$ and $I_j$ respectively through the Bethe ansatz equations.
In the ground state, the two sets of quantum numbers define two 
compact distributions bounded by $J^{\pm}$ and $I^{\pm}$ respectively.
The explicit expressions of $J^{\pm}$ and $I^{\pm}$  are :
\begin{eqnarray}
&J^+-J^-=N_s\qquad\qquad&
J^++J^-=2D_s\nonumber\\
&I^+-I^-=N_c\qquad\qquad&
I^++I^-=2D_c
\label{bound}
\end{eqnarray}
where $N_s$ and $N_c$ have been previously defined while the ``centers"
of the distributions $D_s$ and $D_c$ specify the spin and charge state
respectively. In the ground state, at fixed total momentum $p$,
$D_s$ is the smallest  integer (or half integer) compatible with the
quantization  rules for $J_{\alpha}$. Instead, $D_c=I_h+L/2$ for the Hubbard
model and $D_c=I_h$ for the $t-J$, where $I_h$ defines the position of 
the hole in the distribution of charge rapidities and is related to the
momentum of the holon. Following Woynarovich, we define the
four ``densities" as the $L\to\infty$ limit of $\nu_{c(s)}=N_{c(s)}/L$
and $\delta_{c(s)}=D_{c(s)}/L$, so  that  $D_{c(s)} - \delta_{c(s)} L$ and 
$N_{c(s)} -\nu_{c(s)} $ are finite for $L \to \infty$.
If a hole of momentum $p$ and spin down
is created, the total momentum of the state is given by:
\begin{equation}
-p=-p_0+{2\pi\over L}\left [ (N_s-\nu_s L)(D_s-\delta_s L) +
(N_c-\nu_c L)(D_c-\delta_c L) \right ]
\label{momwoy}
\end{equation}
where $-p_0=2\pi(D_cN_c+D_sN_s)/L$ is the momentum in the thermodynamic limit.
Analogously, the size corrections to the ground state
energy $E$ for the single hole problem can be expressed in terms of  the
above defined quantities, the charge (spin) velocity $v_{c(s)}$, the
elements of the dressed charge matrix $\xi_{ij}$ and of the additional
matrix $Z_{ij}$ as:
\begin{eqnarray}\label{masterwoy}
&&L(E-L\epsilon_{\infty})= - {\pi\over 6}v_s + 2\pi v_c X_c + 2\pi v_s X \\
&&X_c= -(N_c-\nu_cL) \left [ (D_c-\delta_cL)+ \xi_{12}(D_s-\delta_sL)
-\Delta Z_{21}{(N_s-\nu_sL)- \xi_{12}(N_c-\nu_cL) \over 2\xi_{22}}
\right ] \nonumber\\
&&X= \xi_{22}^2 \Big [(D_s-\delta_sL)-Z_{12}(N_c-\nu_cL)\Big ]^2+
\left [{(N_s-\nu_sL)-\xi_{12}(N_c-\nu_cL)\over 2\xi_{22}}\right ]^2
\nonumber
\end{eqnarray}
where $\epsilon_{\infty}$ is the ground state energy per site of the
model at half filling and $N_c-\nu_c L=-1$ for the Hubbard case (where
$N_c=L-1$) and $N_c-\nu_c L=1$ for the $t-J$ model (where $N_c=1$).
The diagonal element of the dressed charge matrix is simply related
to the correlation exponent of the Heisenberg model 
introduced in Section \ref{fieldsec} by
the well known expression $K_\sigma=\xi_{22}^2$ \cite{korepin}.
This general formula is valid both in the Hubbard and in the $t-J$
model. The basic steps for the formal derivation of Eq. (\ref{masterwoy}) 
together with the precise definitions of the quantities appearing in it 
are reported in Appendix \ref{appwoy}. Here we only stress the decoupling 
of the charge and the spin terms in the finite size corrections, in agreement 
with our starting assumptions: The long wavelength Hamiltonian is the sum 
of a charge part which does not give any singularity in the correlation 
functions and a spin part which instead gives rise to critical exponents.
Most importantly, this rather complicate, exact, expression (\ref{masterwoy}) 
perfectly matches the predictions of the bosonization method (\ref{master}).

From Eq. (\ref{momwoy}) we find that the size scaling of the
total momentum is characterized by the amplitude (\ref{momscal})
$\alpha=-(N_s-\nu_s L)(D_s-\delta_s L) -(N_c-\nu_c L)(D_c-\delta_c L)$
while from the magnetic contribution in Eq. (\ref{masterwoy}) 
we easily identify the quantum numbers $(m, n)$ 
and the phase shifts ($\delta,\gamma)$. The other parameters appearing in 
Eq. (\ref{master}) then follow from the charge part of Eq. (\ref{masterwoy})
leading to:
\begin{eqnarray}
\alpha&=&-(N_s-\nu_s L)(D_s-\delta_s L) -(N_c-\nu_c L)(D_c-\delta_c L)
\nonumber\\
m&=&D_s-{1\over 2}\nonumber\\
n&=&(N_s-\nu_sL) \nonumber\\
\delta&=&{1\over 2}-Z_{12}(N_c-\nu_cL) \nonumber\\
\gamma&=&-\xi_{12}(N_c-\nu_cL) \nonumber\\
\omega_+&=&0 \nonumber\\
\omega_-&=& {\Delta Z_{21}(N_c-\nu_cL) \over \xi_{22}}+2Z_{12}(N_c-\nu_cL) 
\end{eqnarray}
where we used that in the ground state $D_s=1/2$ and then $\delta_s=0$.
Note the vanishing of $\omega_+$ which occurs in all the models
we have examined and is probably related to a sort of Luttinger theorem
which forces the volume of the spinon Fermi surface to be unaffected by 
interactions. This exact correspondence between bosonization and Bethe ansatz
demonstrates the validity of our approach in the 
Hubbard and $t-J$ models and allows for the analytical determination
of the phase shifts $\delta_\pm$ governing the singularities of the
one hole Green function.

In general, the coefficients appearing in Eq. (\ref{masterwoy}) are 
non universal quantities which depend on the coupling constants of
the model as well as the average magnetization per site $\mu$ and the holon
momentum. Therefore, we expect that the exponent $X$
is a function of all the parameters which define the hole Hamiltonian,
including the total momentum 
$p$. However, some special but important exception must be mentioned.
At zero magnetic field the spin Hamiltonian possesses the additional $SU(2)$ 
symmetry both in the Hubbard and $t-J$ model. At this particular point, the
exponent $X$ is universal. In fact, the
elements of the matrices $\xi$ and $Z$ acquire analytic values independent
of the total momentum of the state and of the coupling constants:
$\xi_{22}=1/\sqrt 2$, $\xi_{12}=1/2$ and $Z_{12}=0$. Another simple case is 
the $U\to \infty$ limit of the Hubbard model where $Z_{12}=0$ and the dressed 
charge matrix can be expressed in terms of the magnetization per site
$\mu=(N_{\uparrow}-N_{\downarrow})/2L$: $\xi_{12}=1/2-\mu$, while 
$K_\sigma=\xi_{22}^2$
as a function of $\mu$ is shown in Fig. 1 of Ref. \cite{letter}. 
Also in this limit the exponents
do not depend on the total momentum of the state. Finally, when the hole 
sits at the bottom of the band, i.e. if we are at the one hole ground state,
the equations simplify because the holon momentum is always $k_h=\pi$
and so $v_c=0$. Again $Z_{12}=0$ and our expression for the finite size 
corrections coincides with the zero doping {\it limit} of the known form 
valid at finite density. This proves the continuity between the physics of 
the single hole problem and that of finite doping in 1D.

At arbitrary magnetization and momentum no analytical expression for the
phase shifts is available. However, the integral equations reported
in the Appendix \ref{appwoy} can be solved numerically. Few 
examples are reported in Fig. 7 where we show 
the critical exponents $X_m$ (\ref{xs}) as a function of the momentum for 
several magnetizations in the Hubbard model. 
The loci $\omega=\epsilon_h(p+Q_m)$ in the $(p,\omega)$ plane
where the occurrence of the divergence is predicted by our theory
are instead shown in Fig. 8 for a couple of choices of the parameters.
The presence of $2k_F$ zero energy excitations in the spinon spectrum
gives rise to a remarkable symmetry property of these curves:
At a given energy $\omega$, if a singularity occurs at momentum $p$
it will also show up, with possibly different exponent, at momentum $-p+2k_F$.

As a check on the theory of Section \ref{fieldsec} we have numerically
evaluated the overlap of Eq. (\ref{zetadef}) by Lanczos diagonalization in
the $t-J$ model at $J=2t$ in chains up to $32$ sites at magnetization
$\mu=\pm 1/4$. The results are well fitted by a power law behavior in $L$ 
(\ref{zpower}) with an exponent $X$ which clearly depends on the 
momentum of the hole, in agreement with the bosonization analysis. 
A comparison between the numerically determined exponent and the 
prediction of Eq. (\ref{masterwoy}) is shown in Fig. 9 (a). 
An analogous computation has also been performed in the $t-J$ model at
the generic non integrable point $J=t$ and $\mu=1/4$.
The numerical results are also shown in figure 10 (b)
but in this case the comparison with the analytical predictions
based on the Bethe Ansatz solution is not available. However, a quite
similar momentum dependence of the overlap exponent $X$ emerges,
showing that the above features are not special to the exactly integrable
points.

\setcounter{equation}{0}

\section{the shape of the spectral function} 
\label{hubcin}
In this Section we finally discuss the global shape of the spectral function
of one hole in a correlated background. In fact, the bosonization method
developed in 
Section \ref{fieldsec} only concerns the long wavelength, critical 
properties of the Green function and gives no 
information on its short wavelength features.
The exact calculation of a Green function in an interacting system,
at all lengthscales, has been achieved only recently \cite{ha} 
by use of extremely sophisticated methods for the 
Calogero-Sutherland model \cite{calogero}
while no results are available in other interacting systems.
Our task is twofold: to understand the physical nature of the
low lying excitations which contribute to the spectral function and to
develop a useful numerical method for the 
approximate determination of the spectral function $A(p,\omega)$ which can 
be applied to generic one dimensional models. The purpose is to overcome 
the severe finite size effects present in Lanczos diagonalizations 
\cite{horsch,dago95} without resorting to the delicate extrapolations of 
simulation data \cite{preuss} necessary for the computation of
dynamical correlation functions.

Here, we formulate a simple approximation for the one hole Green 
function which captures most of the features of the exact result
and can be usefully applied to interesting correlated models,
in one dimension, like Hubbard. A similar approximation has been
discussed by Penc {\it et al.} \cite{penc}
in the strong coupling limit of the model. For clarity, we introduce the 
method in the framework of the previously discussed $t-J_{XY}$ Hamiltonian
in the $J\to 0$ limit (see Section \ref{txysec}) leaving the study
of the Hubbard model as a final example.

The starting, exact, expression for the 1D Green function of the $t-J_{XY}$
model is Eq. (\ref{green3})
where the spinon function $Z(R,\tau)$ is defined by Eq. (\ref{effe}).
The spinon term $Z$ in fact contains all the interesting correlation
effects as previously pointed out. However, its direct evaluation
proved rather hard even in the simple case of the $XY$ model where the
exact ground state is a Slater determinant, while the more realistic 
case of a Hubbard model cannot be tackled by these methods.
As an approximate way to evaluate the matrix element (\ref{effe})
we can assume that the most relevant contribution to the intermediate states
comes from the {\sl single spinon states}, i.e. from the exact 
eigenstates of $\hat H_\sigma$ in the squeezed chain with only one spinon
excitation. In fact, being $H_\sigma$ an antiferromagnetic
Hamiltonian defined in an odd chain with periodic boundary 
conditions, it gives rise to a frustrated problem. Then, its ground state
contains a free spinon and it is rather natural to assume that a set of
low energy states can be built by giving a finite momentum $Q$ to such a
``quasiparticle". These single spinon states $|Q>$ can be therefore 
labeled by the momentum $Q$, which lies outside the spinon Fermi 
surface, and have energy $\epsilon_\sigma(Q)$ given by the spinon band:
$\epsilon_\sigma(Q)=J \cos Q$ for the $XY$ model, where $\,|Q\,|\,>k_F$
and $k_F=\pi\rho$ is the spinon Fermi momentum.
The single spinon approximation to the function $Z(R,\tau)$ therefore reads:
\begin{equation}
Z(R,\tau)\sim \, {1\over \ell}
\sum_{|Q|>k_F} e^{-i\epsilon_\sigma(Q)\tau} \,e^{iQR}\,
|\,<Q\,|\,(1-n_0)\,|\,\Psi> \,|^2 =
{1\over \ell}\,\sum_{|Q|>k_F} e^{-i\epsilon_\sigma(Q)\tau} \,e^{iQR}\,Z(Q)
\label{ssa}
\end{equation}
where $|\,\Psi>$ is the ground state of the undoped model and 
we have introduced the spinon function $Z(Q)$ 
\begin{equation}
Z(Q)=\sum_R \,e^{-iQR}<\Psi\,|\,  T_\ell^R (1-n_0)\,|\,\Psi>
\end{equation}
This approximation, therefore entirely resides in the assumption that
only the single spinon intermediate states give a finite contribution
to $Z(R,\tau)$. As a consequence, the time dependence of $Z(R,\tau)$ is 
greatly simplified but still not trivial due to the complex structure
which can be present in $Z(Q)$. 

The value of this approximation is that 
the time dependence of $Z(R,\tau)$ is given analytically in terms of the
known spinon excitation spectrum of the model and that only a small
$O(\ell)$ number of matrix elements is necessary for the evaluation
of the full $Z(Q)$. The exact calculation in fact would require the insertion
of a complete set of intermediate states leading to an exponentially 
large number of terms in the sum of Eq. (\ref{ssa}). A consequence of the
single spinon approximation is the presence of a sharp Fermi surface in 
the function $Z(Q)$ which is in fact predicted to vanish
identically for $|Q| < k_F$. This property is well satisfied in the 
quasiparticle weight for the $J\to 0$ limit of the 
$t-J$ model \cite{parola,penc} even if both the numerical calculation of $Z(Q)$ 
in the $t-J_{XY}$ model (see Fig. 5a) and its 
density matrix renormalization group evaluation in the
strong coupling limit of the Hubbard model
show that this is not an exact feature of one 
dimensional systems \cite{qsj}. 

We now proceed to the evaluation of $Z(Q)$ in single spinon approximation
for the exactly soluble $t-J_{XY}$ model, in order to provide a check on
the quality of this approximation. The completeness
condition of the intermediate states which is obeyed by the
exact quasiparticle weight reads:
\begin{equation}
\sum_Q Z(Q)= \ell (1-\rho)
\label{completa}
\end{equation}
In order to compute $Z(Q)$ in the $t-J_{XY}$ 
model on a $L$-site ring and $2\nu+1$ spinons, 
we first recall that the ground state of the $XY$ model is a Slater
determinant of $2\nu+1$ plane waves with momentum quantization (\ref{qj})
and then the occupied orbitals are $|\,q_s>$ with $s\in [-\nu,\nu]$. 
The intermediate states are also Slater determinants of plane waves but with
different quantization rule (\ref{kj}). If the spinon momentum is 
$Q_r={2\pi\over\ell}(r+{1\over 2})$, the first $2\nu$ single particles
intermediate states $|\,k_j>$ fill the spinon Fermi sea $j\in [-\nu,\nu-1]$ 
while the remaining spinon is placed outside this interval, at $j=r$. 
As noted before, the overlap between two Slater determinants is just the 
determinant of the matrix of the overlaps $B_{sj}= <q_s\,|\,k_j>$ 
and the quasiparticle weight $Z(Q_r)$ is the modulus square of such a 
determinant. The numerical evaluation of the
completeness sum rule (\ref{completa}) in single spinon approximation 
is plotted in Fig. 10 for two magnetizations.
The data show that the sum rule is violated in the thermodynamic limit 
and therefore the single spinon states do not represent a complete set
of intermediate states as expected. 
However, Fig. 10 also shows that the breakdown
of the sum rule is very small and shows up at considerably large system size.
This approximation accurately reproduces
the short wavelength properties of the model while fails in catching 
the long wavelength features (i.e. critical exponents and amplitudes) 
which we already discussed by use of bosonization methods in Section 
\ref{fieldsec}. Therefore, we expect that single spinon approximation can
be successfully applied to the study of the global shape of the 
one hole spectral function in one dimensional models. As an example
we now briefly discuss the case of the Hubbard model.

Analogously to the $t-J_{XY}$ model, the Green function is 
expressed in terms of the dynamical quasiparticle weight by Eq. (\ref{green3})
which is approximately evaluated as in (\ref{ssa}): 
\begin{equation}
G(p,\tau)=i\,\sum_Q e^{-i\epsilon_h(p+Q)\tau}
e^{-i\epsilon_\sigma(Q)\tau} |<Q\,|\,c_{p,\downarrow}\, |\,\Psi>|^2
\end{equation}
The remaining problem is
to compute the function $Z(Q) = \ell\,|<Q\,|\,c_{p,\downarrow}\, |\,\Psi>|^2$
for the Hubbard model. 
The numerical calculation can be carried out by Lanczos diagonalizations by
exploiting the negligible size dependence of this quantity already 
verified in the previous examples. The ground state $|\Psi>$ of the 
half filled Hubbard model at magnetization $\mu=-{1\over 2}+{2\nu+1\over L}$
can be numerically obtained in chains up to $L=16$ sites.
The remaining problem is to select the one spinon states $|Q>$
which contribute to $Z(Q)$.
The procedure we have adopted takes advantage of the continuity of the
one spinon states between the weak and the strong coupling limit.
At $U\to\infty$ the one spinon state $|Q>$  is in one to one correspondence 
with the ground state of momentum $Q$ of the Heisenberg chain in a
$\ell=L-1$ site ring. This follows from the factorization property of the 
Hubbard eigenfunctions discussed in Refs. \cite{ogata,parola}. These
spinon states can be therefore identified by performing Lanczos 
diagonalization on the $\ell$ site Heisenberg model in the symmetry subspace
of total momentum $Q$. Now, having determined the one spinon states at
$U=\infty$, we adiabatically lower the interaction parameter 
$U$ (in practice this is quite an easy procedure within Lanczos 
method) following the ``evolution" of the eigenstate of spinon momentum $Q$ 
from $U=\infty$ down to the desired value of $U$. In this way, starting from 
an exact eigenstate of the Heisenberg model we first find an eigenstate
of the Hubbard model at $U=\infty$ and then a sequence of eigenstates of 
the $\ell$ site
Hubbard model corresponding to smaller and smaller interaction parameters.
This procedure has been devised because the spinon momentum $Q$ is a good 
quantum number {\sl only} at $U=\infty$ and then we need a method to select 
the exact one spinon states out of the full set of eigenstates of
the Hubbard Hamiltonian. As usual, a simple check on the validity of the
single spinon approximation comes from the completeness condition of the
intermediate states, analogous to Eq. (\ref{completa}), which now reads
\begin{equation}
{1\over \ell}\,\sum_Q Z(Q) =
<\Psi\, |\,c^\dagger_{p,\downarrow} c_{p,\downarrow}\,|\,\Psi>
\equiv n_\downarrow(p)
\label{completahub}
\end{equation}
i.e. the momentum distribution of the spin down electrons at half filling.
The amount of violation of this sum rule quantifies the weight of all
the other states in the Hilbert space which have been neglected in our
approximation. 

In Figs. 11-12 we show results from Lanczos
diagonalization in chains up to $16$ sites. 
The function $Z(Q)$ for two values of the momentum $p$ of the 
hole and of the magnetization $\mu$ is reported. In all the cases we have
considered the completeness condition (\ref{completahub}) is very well 
satisfied showing that one spinon states account for more than the $98\%$ of
the full Hilbert space of intermediate states \cite{foot}. 
This allows to reconstruct 
the full spectral function for the Hubbard model. On the other hand,
the available data also show weak size dependence 
suggesting that finite size effects are not relevant at high energies.
A plot of the predicted spectral function is shown as a function of
$\omega$ in Fig. 13 for a typical, intermediate coupling 
($U=4t$) and a two hole momenta $p$.

\setcounter{equation}{0}

\section{Summary and discussion}
\label{endsec}
In this work we have analyzed in some detail the dynamical properties of a
hole in an antiferromagnet. Due to the mapping between the attractive
Hubbard model at arbitrary density and the half filled
repulsive Hubbard model in a magnetic field, our analysis
directly applies also to the more general case of hole propagation in
correlated one dimensional
models with a gap either in the charge or in the spin spectrum.
Most of the known quasi one dimensional materials in fact belong to these
classes and then the present study may be helpful in the interpretation
of the available photoemission spectra of quasi $1d$  systems \cite{shen}. 
First we found the exact spectral function of a single hole in the 
Ising model which is characterized by a gap both in the charge and 
in the spin channel. As a result, the quasiparticle weight is finite also
in $d=1$ and $A(p,\omega)$ has a $\delta$ contribution. However, due to the
absence of spin fluctuations the hole dispersion relation is flat and the
hole cannot propagate. Then, we focused 
on the singularities of the hole spectral function which occur because of
the presence of other {\sl gapless} degrees of freedom (spinons). 
This spinon gas behaves as a Luttinger liquid which gives rise to the
typical critical exponents of one dimensional physics that show up
in the hole dynamical properties.
Among the results we have obtained, we like to stress few general features of 
$A(p,\omega)$ which characterize hole propagation in one dimension:
\begin{itemize}

\item{ 1)} The main singularities in the spectral function occur along lines
in the $(p,\omega)$ plane with dispersion relation determined by the 
form of the holon band, $\omega(p)=\epsilon_h(p\pm k_F)$, while the spinon
excitation induced by the hole is created at the Fermi points $\pm k_F$.
This is a consequence of spin charge decoupling which occurs in one
dimension and gives rise to divergences in $A(p,\omega)$ also above the
bottom of the band. In this case, the divergence may occur on both 
sides of $\omega(p)$ with different amplitudes. The existence of 
two branches of singularities in the spectral function (see Fig. 8) can be 
interpreted as due to the presence of a shadow band \cite{penc}.

\item{ 2)} The singularities are characterized by 
critical exponents which can be explicitly calculated in integrable models.
In the isotropic antiferromagnet the $SU(2)$ symmetry forces the exponent
to be exactly $X=1/4$ for all microscopic Hamiltonians. Instead, when spin
isotropy is broken, or when the system has a spin
gap, the critical exponent $X$ in general depends on the parameters
of the model {\sl and} on the momentum of the hole.

\item{ 3)} Away from the $SU(2)$ isotropic point,
the tunneling density of states has either a divergence
or a zero at the bottom of the spectrum according to the value of
the critical exponent $X$, i.e. according to the parameters of 
the model. Remarkably, the density of states shows at most
weak logarithmic singularities in the isotropic case. 

\item{ 4)} At the bottom of the band the critical exponent  
coincides with the known exponent characterizing the spectral weight
in doped systems when the zero doping limit is taken. This provides a
demonstration of the continuity of the physical behavior of the degrees
of freedom which do not develop a gap in the excitation spectrum as 
doping vanishes.

\end{itemize}
All these features should be experimentally detectable in quasi one 
dimensional materials. 

A problem posed by this analysis concerns the relationship between 
our results and the zero doping limit of the generally accepted Luttinger 
liquid picture of the Hubbard or $t-J$ model. In fact, the continuity 
between the single hole and the low doping physics apparently breaks down
when the hole momentum does not coincide with the Fermi momentum of the 
doped model, i.e. when we are above the bottom of the holon band. 
In this case, standard bosonization methods would
predict singularities in the spectral function with momentum independent 
critical exponent uniquely determined by the physics at the Fermi points
\cite{voit} while the accurate analysis of the single hole problem reveals the
presence of momentum dependent critical exponents when we move 
away from the Fermi level. 

From a methodological point of view, this study of the single hole motion
demonstrates a close relationship between the physics of hole motion and
the single impurity problem in Luttinger liquids. This mapping is provided,
at strong coupling, by the Galileo transformation which allows to eliminate
the hole degree of freedom in favor of a non translationally invariant spin 
system. The recoil of the hole, embodied in the hole kinetic contribution 
of the effective spin Hamiltonian $\hat H_p$, 
cuts off the backward scattering
terms in the impurity problem and generates effective boundary conditions 
which allow the propagation of spinons through the impurity site.
This idea can be extended to higher dimensions. The Galileo transformation,
in fact, is not restricted to $d=1$ and the single hole problem can be 
always mapped to a pure spin Hamiltonian. Generalizing what we found here,
it is tempting to assume that also in $d>1$ the hole acts as an effective
boundary condition placed at the origin of the  $d$-dimensional 
spin lattice. The emerging picture resembles and generalizes that of 
the dipolar distortion proposed by Shraiman and Siggia \cite{siggia}
based on the semiclassical treatment of a particular choice of boundary
condition. More work along these directions may eventually clarify the
properties of hole motion in a correlated background in arbitrary 
spatial dimensions.

\vskip 2 true cm
\centerline{\bf ACKNOWLEGMENTS}
\vskip 1 true cm
It is a pleasure to acknowledge useful discussions with M. Fabrizio,
Q. Shaojing, G. Santoro, F. Mila and E. Tosatti. One of us (AP) wishes to thank
the staff at SISSA for warm hospitality. This work is partly supported 
by PRA HTSC of the Istituto Nazionale di Fisica della Materia (SS). 

\vfill\eject

\renewcommand{\theequation}{\thesection.\arabic{equation}}

\setcounter{equation}{0}

\appendix
\section{Formal calculation of the overlap in a Luttinger Liquid}
\label{overapp}
In this appendix we give a formal derivation of the
overlap 
\begin{equation}
O_{\pm} (\delta_{\pm}) =<d_{\pm}|c_{\pm}> 
\end{equation}
between two free particle states
\begin{equation}
<\,d_\pm\,|= \prod_{\mp n\ge 0}\,  d^\dagger_n\,|\,0\,>, \qquad\qquad
<\,c_\pm\,|= \prod_{\mp n\ge 0}\,  c^\dagger_n\,|\,0\,>
\label{defcd}
\end{equation} 
on a given branch of
a Luttinger liquid, here identified by ($\pm$). In order to simplify
the notation we drop the label $\pm$ from the operators $c$ and $d$
whenever it does not lead to ambiguities.
The operators $d$ are defined with skew boundary conditions (i.e. 
with non vanishing phase shifts) while the operators 
$c$ correspond to periodic boundary conditions:
\begin{eqnarray}
c^\dagger_n&=&{1\over \sqrt{\ell}}\int\limits_{-{\ell\over 2}}^{{\ell\over 2}}
e^{ i { 2 \pi \over \ell} n x }
\psi^\dagger_{\pm} (x) \nonumber \\
d^\dagger_n&=&{1\over \sqrt{\ell}}\int\limits_{-{\ell\over 2}}^{{\ell\over 2}}
e^{ i { 2 \pi \over \ell} (n+ \delta ) x }  \psi^\dagger_\pm (x)
\label{formald}
\end{eqnarray}
where $\psi_\pm(x)$ identifies the fermion field in the right $(+)$
or left $(-)$ branch.
The relationship between the operators $d$ and $c$ is easily found using
canonical anticommutation rules for the fields  $\psi_\pm$:
\begin{equation} \label{fonduta}
c^{\dagger}_m= \sum_{n=-\infty}^\infty s_\delta(n-m)  d^{\dagger}_n
\end{equation}
where
\begin{equation}\label{defs}
s_\delta (n)= (-1)^n {\sin( \pi \delta ) \over \pi (n +
 \delta )}
\end{equation}
The overlap $O_\pm(\delta)$ is given by the determinant of the overlap
matrix $D_{n,m}(\delta) =s_\delta (n-m)$ with
the restriction on the allowed indices $n,m \ge 0$ for the left branch and 
$n,m \le 0$ for the right one which selects the occupied orbitals.
From these definitions we get the symmetry property:
\begin{equation} \label{l1}
O_+ (\delta)=O_-(-\delta)
\end{equation}
which follows from the transformation rule of the matrix $D_{n,m} (\delta)$ 
under the mapping $(n,m) \to (-n,-m)$ which changes the left into the right 
branch. A further property of the determinants $O_\pm (\delta)$  
derives from the definition of the matrix $D(\delta)$:
$D_{n,m}(\delta)=s_\delta(n-m)=s_{-\delta}(m-n)=D_{m,n}(-\delta)$
which gives:
\begin{equation} \label{l2}
O_\pm(\delta)=O_\pm(-\delta)
\end{equation}
as the determinant of a matrix is equal to that of its transpose.
In the continuum limit the determinant $O_\pm$ is not well defined and
a cut-off procedure is required before evaluating the overlap.
Some care should be taken in the explicit definition of the cut-off.
In fact, by restricting the matrix indices $(n,m)$ to a finite interval
we would effectively introduce an unphysical doubling of the Fermi surface.
Instead, let us consider a system with both left and right branches with the
same finite phase shifts $\delta_+=\delta_-=\bar\delta$, and finite but large
number of particles $N$ symmetrically distributed in the positive and
negative branch. In this limit it is clear that the left and right branch
decouple and the total overlap $O(\delta)$ is given by the product of the
two left and right component 
$O(\bar\delta)\sim O_+(\bar\delta) O_-(\bar\delta)$.
The overlap $O(\delta)$ can be exactly computed in a finite lattice
with a given (large) number of particles $N$ \cite{parola}
and is formally given by the previously introduced determinant
with matrix indices belonging to the interval $[1,N]$:
\begin{equation} \label{defdeth}
 O(\bar\delta)=  \det { \sin (  \pi \bar \delta   )  \over \pi
(n-m+ \bar \delta) }  = A_{\bar \delta}  N^{- \bar\delta^2 }
\end{equation}
where $A_{\bar \delta}$ is a finite numerical constant.
Then, by use of the relations (\ref{l1},\ref{l2}) and the  previous result
we finally get:
\begin{equation} \label{defodelta}
|\,O_+(\bar\delta)\,|^2 =|\,O_-(\bar\delta)\,|^2 =O(\bar\delta)= 
A_{\bar \delta } N^{-\bar \delta^2 }
\end{equation}
which gives 
$|\,O_\pm (\delta)\, | \, \propto  \ell^{-{1\over 2}\delta^2}$ where we
expressed the number of particles $N$ as a fraction $\rho$ of the number
of sites $\ell$.

\setcounter{equation}{0}

\section{The dynamical quasiparticle weight}
\label{greenapp}
In Eq. (\ref{prodz}) we showed how the dynamical quasiparticle weight $Z(x,t)$
splits into the product of contributions coming from the right and left
fermion branches. Now we have to evaluate the generic matrix element
appearing in the formal expression of $Z(x,t)$ for a non interacting
Fermi gas characterized by given phase shifts $\delta_{\pm}$ at the
two Fermi points:
\begin{equation} \label{defzs}
Z_\pm^\delta (x)= <\,c_\pm\,|\,e^{i x (\hat P_\pm -Q_\pm)}\, |\,c_\pm\,>
\end{equation}
where the states $|c_\pm>$ are defined in appendix A
and  $\hat P_{\pm}$ represents the total momentum operator
for the fermions in the right $+$ or left $-$ branch, defined in 
Eq. (\ref{hfree}), with fermionic operators $\psi_+$ and $\psi_-$ 
obeying skew boundary conditions (\ref{skew}).
Here $Q_+$ is the minimum momentum of the right branch of $d$
electrons (i.e. electrons with a  defined value of the momentum $\hat P$ ),
as excitations in the right $d$ branch can only
increase the momentum by  $ {2 \pi j \over \ell}$, with positive integer $j$.
Conversely,  $Q_-$ is the  maximum allowed momentum in the left branch of
$d$ electrons, as excitations in the left $d$ branch can only decrease 
the momentum by  $ {2 \pi j \over \ell}$ with negative integer $j$.
Though in a Luttinger liquid $Q_\pm$ are infinite constants, the
functions $Z_+(x)$ and $Z_-(x)$ are finite also in the continuum limit
and have an important
property, referred in the following as the Fourier positivity (negativity).
By inserting a complete set of the mentioned excitations with
definite momentum in the RHS of Eq.(\ref{defzs}) we get
\begin{equation} \label{spectz+}
Z_+^{\delta} (x)= \sum_{n\ge 0} \sum_{j}
|\,<\,c_+\,|\,d_{j,n}\,>\,|^2 \,e^{i  { 2 \pi n \over \ell } x }
\end{equation}
where $j$ in $|\,d_{j,n}\,>$ labels all the possible excited states
$|\,d_{j,n}\,>$ with momentum $Q_+ + { 2 \pi n \over \ell }.$
Such a spectral decomposition of $Z_+^{\delta}$ implies  that the Fourier
coefficients $Z_+^\delta (n)= \int_0^\ell
e^{-i {  2 \pi n x \over \ell}  } Z_+^\delta (x) dx$ are non vanishing  
(and positive definite) only for  $ n \ge 0$ (Fourier positivity). 
Analogously $Z_-^\delta (n) > 0$ only for $n\le 0$ (Fourier negativity ).

In terms of the $d$ operators (\ref{formald}) the momentum on each branch
$\hat P_\pm$ is diagonal and reads:
\begin{equation}\label{defps}
\hat P_\pm = \sum_n  (k_n+k_F^\pm) d^{\dagger}_n d_n
\end{equation}
where $k_n= { 2 \pi \over \ell } (n+\delta)$
and $k_F^\pm=\pm k_F$.
Using the same particle hole transformation discussed in appendix A,
(i.e. the transformation $n\to -n$) we obtain the analog of Eq.(\ref{l1}):
\begin{equation} \label{ll1}
Z_{\pm}^{\delta} (x)=Z_{\mp}^{-\delta} (-x)
\end{equation}

Having discussed the general symmetry properties of $Z^\delta_\pm(x)$, 
we now turn to the explicit evaluation of the function 
\begin{equation}
Z^\delta(x)=<\,\Psi\,|e^{i\hat P x}\,|\,\Psi\,>
\end{equation}
in a free Fermi gas of $N$ particles, where $|\,\Psi\,>$ is the ground 
state with 
periodic boundary conditions while the momentum operator $\hat P$
refers to a system with {\sl skew} boundary conditions. Following appendix 
\ref{overapp}, we carry out the calculation for a model of fermions 
with constant phase shift $\delta$ throughout the Brillouin zone
and then we relate the result to $Z^\delta_\pm(x)$ by use of the 
symmetry properties previously discussed. In particular, if we keep only
the most relevant singularity, the factorization property proved in Eq. 
(\ref{prodz}) gives:
\begin{equation}
Z^\delta(x)=e^{iQx} \,Z^\delta_+(x)\,Z^\delta_-(x)
=e^{iQx} \,Z^\delta_+(x)\,Z^{-\delta}_+(-x)
\label{fundz}
\end{equation}
where use has been made of the symmetry (\ref{ll1}) and $Q=Q_++Q_-$ 
is the reference momentum of the intermediate states.

In a free Fermi gas, both the ground state $|\,\Psi\,>$ and the intermediate
states are Slater determinants built with the different fermionic operators
$c_n$ and $d_n$ respectively (see Eqs. \ref{defcd} and \ref{formald}).
The translation operator $\exp (i\hat P x)$ with
$\hat P=\hat P_++\hat P_-$ (\ref{defps}) is a one body unitary operator
which maps the Slater determinant with plane waves single particle orbitals 
\begin{equation}
\phi_n(r)={1\over\sqrt{\ell}} \,e^{i {2\pi\over \ell} nr}
\end{equation}
into another Slater determinant with orbitals 
\begin{equation}
\psi_m(r)={1\over\sqrt{\ell}} \,\sum_j\,s_\delta(j-m)\,e^{i\,k_j(x+r)}
\end{equation}
where $s_\delta(n)$ is defined in (\ref{defs}).
The overlap between these two Slater determinants is just the 
determinant of the $N\times N$ matrix $M_{nm}$ of the overlaps
between the occupied orbitals:
\begin{equation}
M_{nm}=z^\delta\,\sum_j\,s_\delta(j-m)\,s_\delta(j-n)\,z^j
\end{equation}
giving explicitly
\begin{equation}
M_{nm}=\cases{ 
{\displaystyle 
(-1)^{n-m}\,e^{i\pi\delta}\,
{\sin\pi\delta\over\pi}\,{z^n-z^m\over n-m}}
& for $n\ne m$ \cr
{\displaystyle 
z^n\left [ 1+iue^{i\pi\delta}\,{\sin\pi\delta\over\pi}\right ]}
& for $n=m$ \cr}
\end{equation}
where we have introduced the phase factor $z=e^{i u}$ with 
$u={2\pi x\over\ell}$. The matrix $M_{nm}$ can be 
written as the product three diagonal matrices (with diagonal elements
$(-1)^n$, $z^n$ and $(-1)^m$) which contribute to the determinant
with a phase factor and the Toeplitz matrix $T_{n-m}$ given by:
\begin{eqnarray} \label{toepliz}
T_n &=& \int_0^{2 \pi} {d\theta\over 2 \pi} C(\theta) e^{-i \theta n }
\nonumber \\
C(\theta) &=& 1 +   \Theta ( u -\theta)  (e^{ i 2\pi\delta } -1 )
\end{eqnarray}
where $\Theta$ is the step function. The leading singularity of the 
determinant of $T$ might be extracted by means of Szeg\"o's 
theorem \cite{mccoy} which would give:
\begin{equation}
\det T=\exp\left \{ N g_0 + \sum_{n \ge 1 } n g_{-n} g_n \right \}
\end{equation}
with 
\begin{equation}
g_n=\int_0^{2 \pi } { d\theta\over 2 \pi } e^{-in\theta}\,\ln C(\theta) 
\end{equation}
By performing the Fourier transform, $g_n$ is simply evaluated:
\begin{equation} \label{defgn}
\cases{g_0=  i \delta\,\,u & \cr
g_n={\delta \over n } ( 1-e^{-i u n } ) & for $n\ne 0$ \cr }
\end{equation}
However, this theorem holds only for continuous $C(\theta)$ 
and does not apply directly
to our case. However one can follow the same regularization applied in
an analogous calculation for the Ising model \cite{mccoy} by noting
that the Hilbert matrix of elements $H_{n-m}$ defined by:
\begin{equation}
H_n={\sin\pi\delta\over \pi(n+\delta)}
\label{hilbert}
\end{equation}
has the same kind of singularity shown by $T$. In fact, the coefficients 
$\bar g_n$ which characterize $H$ are just given by $g_n=\delta/n$ for
$n\ne 0$ and $g_0=0$. Therefore, following Ref. \cite{mccoy}, we can
apply Szeg\"o's theorem to the ratio of determinants
\begin{equation}
{\det T\over (\det H)^2}=
\exp\left \{ N g_0 + \sum_{n \ge 1 } n \left [g_{-n} g_n +{2\delta^2\over n^2}
\right]\right \}
\end{equation}
Finally, the determinant of a Hilbert matrix can be analytically evaluated,
as shown in Eq. (\ref{defdeth}), giving, to leading order,
\begin{equation}
Z^\delta(x)=\det T = \, e^{iQx}\,g^+(x)\,g^+(-x) 
\label{ftp}
\end{equation}
where the overall phase factor depends on the reference momentum of the 
intermediate states $Q=\sum_n k_n $ (for $k_n<k_F$) and 
\begin{eqnarray}
g^+(x)&=&A_\delta\,N^{-\delta^2} 
\exp\left[ \delta^2 \sum_{n \ge 1}  { e^{ i u n }  \over n} \right ] 
\label{g+}\\
&=&A_{\delta } \left[ N (1-e^{ i r})\right]^{-\delta^2 }
\label{findet}
\end{eqnarray}
$A_\delta$ is the same numerical constant appearing in Eq. 
(\ref{defdeth}) and is given explicitly by
\begin{equation} \label{defa}
\ln A_{\delta} = - (\delta )^2 ( 1 + C) + \sum_{j=1}^\infty j \left[
{\delta ^2 \over j^2 } +  \ln(1 -{\delta ^2 \over j^2} ) \right]
\end{equation}
$C=0.5772 \ldots$ being the Euler constant.

The Fourier positivity property of $g^+(x)$ can be easily proved by expanding 
the exponential in Eq. (\ref{g+}). 
By comparing Eq. (\ref{ftp}) and Eq.(\ref{fundz}) we find:
\begin{equation}
g^+ (x)/Z_+^\delta (x)=g^+(-x)/Z_+^{-\delta} (-x)
\end{equation}
where the left hand side has the Fourier positivity property
(as the ratio of two function satisfying this property also satisfies the
Fourier positivity property),  while the right hand side has the Fourier
negativity property. Therefore both terms of this equation have to  be
constant in $x$ implying  that $Z_+^{\delta} (x)  \propto g^+(x)$.
The overall proportionality constant can be determined by noting that
the $n=0$ Fourier coefficient in Eq. (\ref{spectz+})  
coincides with the square of the overlap $|O_+(\delta)|^2$ which has been
explicitly calculated in Eq.(\ref{defodelta}). On the other hand, the
$n=0$ Fourier component of $g^+(x)$ can be read off from Eq. (\ref{g+})
yielding $Z_+^{\delta}(x)=g^+(x)$.

In the thermodynamic limit ($\ell\to \infty$ at fixed $x$ and fixed
density of fermions $\rho={N\over L}$) Eq. (\ref{findet}) simplifies:
\begin{equation} \label{finalz}
Z^\delta_+(x)=A_{\delta } \,(2 \pi \rho)^{-\delta^2} 
\,(-i x +\epsilon)^{-\delta^2 }
\end{equation}
with $\epsilon\sim O(x^2 /\ell^2)$ is a vanishingly small positive term
which defines the branch cut for the non integral exponentiation of 
the complex number $(-ix+\epsilon)$. Taking the Fourier transform, we
get (for $0<\delta<1 $):
\begin{equation} \label{ftz+}
Z_\pm^{\delta}(k)=  (2 \pi \rho)^{-\delta^2} \,
A_{\delta}  2 \sin (\pi \delta^2)\Gamma(1 -\delta^2)  \Theta(\pm k)
(\pm k)^{\delta^2 -1} 
\end{equation}

The asymptotic evaluation of the two functions $Z_\pm^\delta(x)$ 
for a Fermi gas with arbitrary phase shift also provides the 
leading singularity for 
$Z_\pm^{(\delta_\pm,\pm m)}(x)=Z_\pm^{(\delta_\pm+m)}(x)$ 
entering the spectral function of the interacting
model with different phase shifts on the two branches $\delta_\pm$.
In particular, for $t=0$, the Fourier transform of $Z(x,0)$
in Eq. (\ref{prodz}) splits into the sum of terms with different phase shifts 
$\bar\delta_+=\delta_++m$ and $\bar\delta_-=\delta_-+m$ each given by
\begin{eqnarray} \label{zqgen}
Z(Q+k) &= B  \sin (\pi \bar \delta_+^2 )
 k^{  \bar \delta_+^2 +\bar \delta_-^2 -1}~~~~~~&{\rm for ~ k> 0}  \nonumber \\
Z(Q+k) &= B  \sin (\pi \bar \delta_-^2 )
 (-k)^{  \bar \delta_+^2 +\bar \delta_-^2 -1}~~&{\rm for ~ k<0}
\end{eqnarray}
if $Q$ is the average spinon momentum. The prefactor can be also calculated:
$B= 2 (2 \pi \rho)^{-(\bar\delta_+^2+\bar\delta_-^2)}\, 
A_{\bar \delta_+} A_{\bar \delta_-} 
\Gamma (1 - \bar\delta_+^2 -\bar \delta_-^2) $.
In this case, the Fourier transform is nonvanishing both for positive and
negative $k$ and shows singularities with the same exponent for $k\to 0^\pm$.
The amplitude ratio tends to a number which, in the free Fermi gas, 
just depends on the phase shifts:
\begin{equation} \label{a+/a-}
{A_+\over A_-} =\lim\limits_{k \to 0^+}
{Z(Q+k)\over Z(Q-k)}={ \sin (\pi \bar \delta_+^2) \over
\sin (\pi \bar \delta_-^2) }
\end{equation}
The asymmetry between the two sides of the singularity is therefore 
enhanced when one of the phase shifts gets small. 

\setcounter{equation}{0}

\section{the long wavelength expansion}
\label{aptxy}
The first task is to evaluate the summation present in Eq.
(\ref{overmat}) at large $R$. For simplicity we restrict the analysis 
to the low density regime. It is easy to verify that the most relevant
contributions to the sum are those around $k_j\sim 0$. For $r\ne s$
this gives the approximate form:
\begin{equation}
A_{rs}\sim {4 e^{iJ\tau} \over L (q_r-q_s)}\left [ f(q_r)-f(q_s)\right ]
\label{ars}
\end{equation}
where the function $f(q)$ is given by:
\begin{equation}
f(q)=\int_{-\pi}^{\pi} {dk\over 2\pi} {e^{i(kR-\tau J k^2/2)}\over k-q}
\label{effino}
\end{equation}
Standard asymptotic expansion leads to
\begin{equation}
f(q)=e^{i(qR-\tau J q^2/2)} \,i\, \int_{-\infty}^{\infty} {dx\over 2\pi}
e^{-i\alpha x^2}\,{\sin x\over x}
\label{effino2}
\end{equation}
where $\alpha= [z/(1+qz)^2]/(2R)$ and $z=-J \tau/R$. In the $R\to \infty$ limit
at $z=$const, $\alpha\to 0 $ and the problem simplifies. Notice that
this last limiting procedure requires that $\alpha$ is regular and therefore
that $1+qz$ is always positive. This implies $|z| < 1/k_F$ (where $k_F$ is
the Fermi momentum of the spinon) which is always satisfied at low density.
In general this inequality leads to $R>v_s\tau$ where $v_s=J k_F$ is the spinon
velocity at low density. In the following we will consider only this
regime. In this case, $\alpha$ can be set equal to zero in the integral
leading to the final expression:
\begin{equation}
f(q)={i\over 2}\,e^{i(qR-\tau J q^2/2)}
\label{effino3}
\end{equation}
which inserted into (\ref{ars}) gives
\begin{equation}
A_{rs}\sim {e^{iJ\tau } i \over \pi (r-s)}\left [
e^{i(x r + y r^2)}-e^{i(x s + y s^2)}
\right ]
\label{ars1}
\end{equation}
where $r$ and $s$ run over the occupied spinless fermion orbitals $[-\nu,\nu]$
and $x=2\pi R/L$, $y=-2\pi^2 J\tau /L^2$. The diagonal elements require
a separate analysis which gives
\begin{equation}
A_{rr}\sim e^{iJ\tau}  e^{i(x r + y r^2)}
\left [ 1-{x+2y r\over \pi} \right ]
\label{ars2}
\end{equation}
Now it is convenient to express the matrix $A_{rs}$ as the product of
a real matrix $B_{rs}$ and other diagonal matrices
$D_{rr}=e^{i(x r + y r^2)/2}$. In fact, $A= D\,B\,D$ where
\begin{eqnarray}
B_{rs}&\sim& -{2\sin \left [{1\over 2}(x (r-s) + y (r^2-s^2))
\right ]\over \pi (r-s)} \qquad\quad r\ne s \nonumber\\
B_{rr}&\sim& 1-{x +2y r \over \pi}
\label{brs}
\end{eqnarray}
The asymptotic form of the $N$-spinon Green function $Z(R,\tau)$
is then given, besides a global phase factor coming from the determinant of $D$,
by the determinant of the real $N\times N$ matrix $B_{rs}$.
For convenience, we fix a finite ratio $N/L=\rho$ (even if the expressions
previously derived are exact only for $\rho\to 0$) and we perform
the numerical computation of $\det B$ for several values of $R$ and $\tau$.
The results are reported in Fig. 6

\setcounter{equation}{0}

\section{Derivation of the finite size corrections}
\label{appwoy}

In this Appendix we sketch the derivation of the general formula 
(\ref{masterwoy}) for the finite size corrections in the Bethe ansatz soluble
models: Hubbard and $t-J$ at $J=2t$. 
We closely follow the procedure detailed by Woynarovich in 
his work on the finite size corrections of the Hubbard model
at finite doping \cite{woy}. For sake of clarity, Hubbard and $t-J$ models 
will be treated separately.

\subsection{Hubbard Model}

We consider the Hubbard Hamiltonian on a $L$ site chain 
at fixed chemical potential $\mu$ and magnetic field $h$
(in units of the hopping amplitude $t$):

\begin{equation}
H=-\sum_{i=1}^L \sum_{\sigma} ( c^{\dagger}_{i+1,\sigma}c_{i,\sigma} +
c^{\dagger}_{i,\sigma}c_{i+1,\sigma}) + U\sum_{i=1}^L 
n_{i,\uparrow}n_{i,\downarrow} +\mu\sum_{i=1}^L(n_{i,\uparrow}+
n_{i,\downarrow})-{1\over 2}h\sum_{i=1}^L(n_{i,\uparrow}-n_{i,\downarrow})
\end{equation}
where $\sigma=\uparrow,\,\downarrow$ is the electron spin index.
The Bethe ansatz equations for the Hubbard chain read:
\begin{eqnarray}
&&Lk_j=2\pi I_j +\sum_{\beta=1}^{N_s}
2\,\arctan\left(4\,{\sin k_j-\lambda_{\beta}
\over U}\right )\nonumber\\
&&\sum_{j=1}^{N_c}2\,\arctan\left(4\,{\lambda_{\alpha} - \sin k_j \over U}
\right )=2\pi J_{\alpha}+\sum_{\beta=1}^{N_s} 
2\,\arctan\left(2\,{\lambda_{\alpha}-\lambda_{\beta}\over U}\right )
\label{bethe}
\end{eqnarray}
where $N_c=L-1$ and $N_s=N_{\downarrow}$. The quantum numbers $I_j$ and
$J_{\alpha}$ are integers or half odd integers depending on the 
parities of $N_c$ and $N_s$: $I_j=N_s/2\,{\rm mod}(1)$, 
$J_{\alpha}=(N_c+N_s+1)/2\,{\rm mod}(1)$. The existence of a solution to
these equations requires that each set $I_j$ and $J_{\alpha}$ 
consists of mutually different quantum numbers.
Therefore, the distribution $I_j$ is uniquely defined by the position
of the hole $I_h$.
Due to the periodicity of the Bethe ansatz equations by the substitution
$I_j\to I_j+L$ and $k_j\to k_j+2\pi$ we can always assume that 
the $L-1$ quantum numbers $I_j$ fill the range $[I_h+1,I_h+L-1]$.
The important low energy 
real solutions are found if the $N_s$ quantum numbers $J_\alpha$ 
are chosen as contiguous integers (or half odd integers). We denote as
$J_{\rm min}$ and $J_{\rm max}$ the minimum and maximum value of the 
distribution respectively. The 
ground state corresponds to the most symmetrical distribution 
around zero compatible with the quantization rules. 
It is useful to introduce the additional quantities $I^-=I_h-1/2$,
$I^+=I_h+L-1/2$, $J^-=J_{\rm min}-1/2$ and $J^+=J_{\rm max}+1/2$
which, by definition, satisfy the relations (\ref{bound}).
Finally, the total energy is 
expressed in terms of the rapidities $k_j$ and $\lambda_{\alpha}$ by
\begin{equation}
E=-2\sum_j^{N_c}\cos k_j +\mu N_c+h(N_s-N_c/2)
\end{equation}

The Bethe ansatz equations (\ref{bethe}) can be written as
$z_c(k_j)=I_j/L$ and $z_s(\lambda_{\alpha})=J_{\alpha}/L$
in terms of the functions $z_c(k)$ and $z_s(\lambda)$ defined
by Eqs. (2.6) of Ref. \cite{woy}. Following this work we also
introduce the boundaries $k^{\pm}$ and $\lambda^{\pm}$ of the rapidity 
distributions defined by 
$z_c(k^{\pm})=I^{\pm}/L$ and $z_s(\lambda^{\pm})=J^{\pm}/L$.
The distribution function for the rapidity $\rho_c(k)$  
is defined as the derivative of $z_c(k)$ with respect to $k$
and, due to Eq. (\ref{bound}), satisfies
\begin{eqnarray}
&&\phantom{-{1\over 2}}
\int_{k^-}^{k^+}\rho_c(k)={N_c\over L} \nonumber \\
&&-{1\over 2}\left (\int_{k^-}^{2\pi}\rho_c(k)-\int_{2\pi}^{k^+}
\rho_c(k)\right )
-{1\over 2\pi}\int_{\lambda^-}^{\lambda^+}2\,\arctan(4\lambda/U)
\rho_s(\lambda)={D_c\over L}-1 \nonumber \\
\label{a1}
\end{eqnarray}
while the analogous distribution function $\rho_s(\lambda)$ obeys Eqs. (2.14)
of Ref. \cite{woy}.
Following the derivation of Ref. \cite{woy} we find the
$O(1/L^2)$ correction to the rapidity distributions which can
be expressed in compact notation as:
\begin{eqnarray}
\rho(k,\lambda)&=&\rho_{\infty}(k,\lambda)+ \\
&\phantom{=}&{1\over 24 L^2}
\left ({\rho_1(k,\lambda | k^+\lambda^+)\over\rho_c(k^+)} -
{\rho_1(k,\lambda | k^-\lambda^-)\over\rho_c(k^-)} +
{\rho_2(k,\lambda | k^+\lambda^+)\over\rho_s(\lambda^+)} -
{\rho_2(k,\lambda | k^-\lambda^-)\over\rho_s(\lambda^-)}\right )
\nonumber
\end{eqnarray}
where $\rho$, $\rho_1$, $\rho_2$ are vector functions with two 
components, one referring to charge and the other to spin, and
satisfy equations (2.16),(2.20) of Ref. \cite{woy}.
Notice that in our one hole case the functions $\rho_1$ are symmetric
under the simultaneous interchange of $k^+\to k^-$ and 
$\lambda^+\to \lambda^-$ while the functions $\rho_2$ are antisymmetric.
This property leads to the cancellation of the charge contribution to 
the finite size corrections of the total energy which now reads 
\begin{equation}
E=L\epsilon_{\infty}(k^+,k^-,\lambda^+,\lambda^-) - 
{1\over 12 L} \epsilon_2(k^+,k^-,\lambda^+,\lambda^-)
\label{etot}
\end{equation}
where $\epsilon_{\infty}$ and $\epsilon_2$ are defined by Eqs. (2.23)-(2.25)
of Ref. \cite{woy}. In the thermodynamic limit $k^-\to k_0$
(i.e. the holon momentum) and $k^+\to k_0+2\pi$ while, as usual, the
spin rapidities $\lambda$ are centered symmetrically around the
origin: $\lambda^+\to \lambda_0$ and $\lambda^-\to -\lambda^0$.
By expanding $\epsilon_{\infty}(k^+,k^-,\lambda^+,\lambda^-)$ 
around its limiting value for infinite size we get:
\begin{eqnarray}
\epsilon_{\infty}(k^+,k^-,\lambda^+,\lambda^-) &=& 
\epsilon_{\infty}(k_0+2\pi,k_0,\lambda_0,-\lambda_0) +\nonumber \\
&\phantom{=}&
\pi v_c(k_0) \rho_{\infty c}(k_0) 
\left [ (k^+-k_0-2\pi)^2 - (k^--k_0)^2\right] +
\nonumber\\
&\phantom{=}&
\pi v_s(\lambda_0) \rho_{\infty s}(\lambda_0) 
\left [ (\lambda^+-\lambda_0)^2 + (\lambda^-+\lambda_0)^2\right]
\label{epsinf}
\end{eqnarray}
in terms of the charge ($v_c(k_0)$) and spin ($v_s(\lambda_0)$) velocities.
The next step is to express the difference between $\lambda^{\pm}\,$ 
[$k^{\pm}$] and its asymptotic value $\lambda_0\,$ [$k_0$]
in terms of the known parameters $N_{c(s)}$
and $D_{c(s)}$. To this end we start from the Eqs. (\ref{a1}) and 
evaluate their derivatives with respect to $k^{\pm}$ and $\lambda^{\pm}$
in the thermodynamic limit. The final expressions for $\nu_c$ and $\nu_s$
coincide with Eqs. (2.34) of Ref. \cite{woy} in which the $k$ integration
is extended to the full interval $[0,2\pi]$.
Instead, the equations for $\delta_c$ and $\delta_s$ now read:
\begin{eqnarray}
{\partial \delta_c\over \partial k^+}&=&-{\partial\delta_c\over \partial k^-}
+\rho_{\infty c}(k_0)\nonumber\\
&=& -{\rho_{\infty c}(k_0)\over 2}
\left( -1+\int_{k_0}^{2\pi}\sigma_{1c}(k)
-\int_{0}^{k_0}\sigma_{1c}(k) \right )-{\rho_{\infty c}(k_0)\over \pi}
\int_{-\lambda_0}^{\lambda_0}\arctan(4\lambda/U)\sigma_{1s}(\lambda)
\nonumber\\
&\equiv &\rho_{\infty c}(k_0)Z_{11} \nonumber\\
{\partial\delta_s\over \partial k^+}&=&-{\partial\delta_s\over \partial k^-}=
-{\rho_{\infty c}(k_0)\over 2} \left ( 
\int_{\lambda_0}^{\infty}\sigma_{1s}(\lambda) -
\int_{-\infty}^{-\lambda_0}\sigma_{1s}(\lambda) \right )
\equiv \rho_{\infty c} (k_0)Z_{12} \nonumber\\
{\partial\delta_c\over \partial \lambda^+}&=&
-{\rho_{\infty s}(\lambda_0) \over 2} \left (
\int_{k_0}^{2\pi}\sigma_{2c}(k) -\int_{0}^{k_0}\sigma_{2c}(k) 
\right ) -\nonumber\\
&\phantom{=}& {\rho_{\infty s}(\lambda_0) \over \pi} \arctan(4\lambda_0/U)
-{\rho_{\infty s}(\lambda_0) \over \pi} \int_{-\lambda_0}^{\lambda_0}
\arctan(4\lambda/U)\sigma_{2s}(\lambda) 
\equiv \rho_{\infty s}(\lambda_0)Z_{21}^+ \nonumber\\
{\partial\delta_c\over \partial \lambda^-}&=&
{\rho_{\infty s}(\lambda_0) \over 2} \left (
\int_{-2\pi}^{-k_0}\sigma_{2c}(k) -\int_{-k_0}^0\sigma_{2c}(k) 
\right )-\nonumber \\
&\phantom{=}& {\rho_{\infty s}(\lambda_0) \over \pi} \arctan(4\lambda_0/U)
-{\rho_{\infty s}(\lambda_0) \over \pi} \int_{-\lambda_0}^{\lambda_0}
\arctan(4\lambda/U)\sigma_{2s}(\lambda) 
\equiv \rho_{\infty s}(\lambda_0)Z_{21}^- \nonumber\\
{\partial\delta_s\over \partial \lambda^+}&=&{\partial\delta_s\over 
\partial \lambda^-}=-{\rho_{\infty s} (\lambda_0)\over 2} \left( -1+
\int_{\lambda_0}^{\infty}\sigma_{2s}(\lambda)
-\int_{-\infty}^{-\lambda_0}\sigma_{2s}(\lambda)
\right )\equiv \rho_{\infty s}(\lambda_0)Z_{22} 
\label{monster}
\end{eqnarray}
where the additional functions 
\begin{eqnarray}
&&\sigma_{1c(s)}={\partial\rho_{c(s)}\over \partial k^+} \nonumber \\
&&\sigma_{2c(s)}={\partial\rho_{c(s)}\over \partial \lambda^+} 
\end{eqnarray}
satisfy the equations (2.16), (2.36) of Ref. \cite{woy}. 
From their definitions, it is easy to see that the equations
satisfied by the elements of the dressed matrix simplify and give
$\xi_{11}=1$ and $\xi_{21}=0$ while 
\begin{eqnarray}
\xi_{22}(\lambda) &=&1
- \int_{-\lambda_0}^{\lambda_0} {d\lambda'\over 2\pi} 
K_2(\lambda-\lambda') \xi_{22}(\lambda') \nonumber \\
\xi_{12}(k)&=&
\int_{-\lambda_0}^{\lambda_0} {d\lambda\over 2\pi} 
K_1(\sin k-\lambda) \xi_{22}(\lambda) 
\label{charges}
\end{eqnarray}
where the kernels $K_1(x)$ and $K_2(x)$ are defined by:
$K_1(x)=8U/(U^2+16 x^2)$ and $K_2(x)=4U/(U^2+4x^2)$. Some of the elements
of the additional matrix $Z_{ij}$ can be related to the 
dressed charges $\xi_{ij}$ by the algebraic relations:
\begin{eqnarray}
Z_{11}&=&{1\over 2}-\xi_{12}(k_0)Z_{12} \nonumber\\
Z_{22}&=&[2\xi_{22}(\lambda_0)]^{-1} \nonumber\\
Z_{21}^++Z_{21}^-&=&-\xi_{12}(k_0)/\xi_{22}(\lambda_0) 
\label{zxi}
\end{eqnarray}
while the remaining combinations are expressed in terms
of $\sigma_{1s}$ and $\sigma_{2s}$ by
\begin{eqnarray}
Z_{12}&=&-{1\over 2}\left [
\int_{\lambda_0}^{\infty} d\lambda \,\sigma_{1s}(\lambda)-
\int_{-\infty}^{-\lambda_0} d\lambda \,\sigma_{1s}(\lambda)\right ]
\nonumber \\
\Delta Z_{21}\equiv Z_{21}^+-Z_{21}^-&=&\int_{-\sin k_0}^{\sin k_0} dt \left [
{1\over 2\pi} K_1(t-\lambda_0) +
\int_{-\lambda_0}^{\lambda_0} {d\lambda\over 2\pi} K_1(t-\lambda)
\sigma_{2s}(\lambda) \right ]
\label{zeta}
\end{eqnarray}
The two functions $\sigma_{1s}(\lambda)$ and $\sigma_{2s}(\lambda)$ 
satisfy the following equations:
\begin{eqnarray}
\sigma_{1s}(\lambda)&=&{1\over 2\pi} K_1(\lambda-\sin k_0)
- \int_{-\lambda_0}^{\lambda_0} {d\lambda'\over 2\pi} 
K_2(\lambda-\lambda') \sigma_{1s}(\lambda') \nonumber \\
\sigma_{2s}(\lambda)&=&-{1\over 2\pi} K_2(\lambda-\lambda_0)
- \int_{-\lambda_0}^{\lambda_0} {d\lambda'\over 2\pi} 
K_2(\lambda-\lambda') \sigma_{2s}(\lambda') 
\label{sigma}
\end{eqnarray}
By substituting Eq. (\ref{epsinf}) into Eq. (\ref{etot}) and
evaluating the quantities in brackets by means of Eq. (\ref{monster})
we get Eq. (\ref{masterwoy}). 

In conclusion, the relevant equations 
for the one hole problem in the Hubbard model are:
\begin{eqnarray}
\epsilon_s(\lambda)&=&h_s-2\int_{-\pi}^{\pi} {dk\over 2\pi} K_1(\lambda-\sin k)
\cos^2 k 
- \int_{-\lambda_0}^{\lambda_0} {d\lambda'\over 2\pi} 
K_2(\lambda-\lambda') \epsilon_s(\lambda') \nonumber \\
\epsilon_c(k)&=&h_c - 2\cos k 
+ \int_{-\lambda_0}^{\lambda_0} {d\lambda\over 2\pi} 
K_1(\sin k-\lambda) \epsilon_s(\lambda) 
\label{bands}
\end{eqnarray}
which give the spin and charge excitation energies. Obviously, the holon energy
is $\epsilon_h=-\epsilon_c$.  The external fields
$h_s$ and $h_c$ are chosen in such a way that $\epsilon(\lambda_0)=0$
and $\epsilon(k_0)=0$. The charge and spin distributions are given by:
\begin{eqnarray}
\rho_s(\lambda) &=& \int_{-\pi}^{\pi} {dk\over 4 \pi^2} K_1(\lambda-\sin k)
- \int_{-\lambda_0}^{\lambda_0} {d\lambda'\over 2\pi} 
K_2(\lambda-\lambda') \rho_s(\lambda') \nonumber \\
\rho_c(k)&=&{1\over 2\pi}+\cos k
\int_{-\lambda_0}^{\lambda_0} {d\lambda\over 2\pi} 
K_1(\sin k-\lambda) \rho_s(\lambda) 
\label{rho}
\end{eqnarray}
The spinon and holon velocities are
\begin{eqnarray}
2\pi v_s&=&{1\over \rho_s(\lambda_0)}{d \epsilon_s(\lambda)\over d\lambda }
\Big\vert_{\lambda_0}
\nonumber \\
2\pi v_c&=&-{1\over \rho_c(k_0)}{d \epsilon_c(k)\over d k }
\Big\vert_{k_0}
\label{velocity}
\end{eqnarray}
The cut--off $\lambda_0$ and the holon (dressed) momentum $k_0$ are
related to the magnetization and to the position of the hole in the 
charge distribution by
\begin{eqnarray}
{I_h\over L}&=&\int_{0}^{k_0} dk \rho_c(k)\nonumber\\
{N_s\over L}&=&\int_{-\lambda_0}^{\lambda_0} d\lambda \rho_s(\lambda)
\label{mag}
\end{eqnarray}
Finally, the elements of the dressed charge matrix (evaluated at the
cutoff $k_0$ and $\lambda_0$) are given by Eqs. (\ref{charges}) and 
the additional matrix is defined by Eqs. (\ref{zeta}) with the the help of
Eqs. (\ref{sigma}). The finite size corrections to the energy and the
correlation exponent $X_s$ are then given by Eq. (\ref{masterwoy}) in terms of
the above defined quantities.

\subsection{Supersymmetric t-J Model}

The expressions for the $t-J$ model are quite similar. In particular the 
formal size corrections (\ref{masterwoy}) are the same while only the definition
of the coefficients are different. Also the procedure closely follows that
outlined for the Hubbard model. Therefore here we will just report the final
expressions in Sutherland's representation \cite{suth,bares} noting that
in this formalism there is only one ``charge" (i.e. the hole): $N_c=1$.

The spin and charge excitation energies are given by the equations
\begin{eqnarray}
\epsilon_s(v)&=&h_s-2\dot\theta(2v) 
- \int_{-v_0}^{v_0} {dv'\over 2\pi} 
\dot\theta(v-v') \epsilon_s(v') \nonumber \\
\epsilon_c(w)&=&h_c 
+ \int_{-v_0}^{v_0} {dv\over 2\pi} 2\dot\theta(2v-2w) \epsilon_s(v) 
\label{bandstj}
\end{eqnarray}
where $h_s$ and $h_c$ are chosen in such a way $\epsilon(v_0)=0$
and $\epsilon(w_0)=0$. 
The kernel $\theta(x)=2\arctan x$ is defined according to Bares {\it et al}
\cite{bares}. Dot represents derivation with respect to the argument. 
The charge and spin distributions are given by:
\begin{eqnarray}
\rho_s(v) &=& {1\over \pi}\dot\theta(2v)
- \int_{-v_0}^{v_0} {dv'\over 2\pi} 
\dot\theta(v-v') \rho_s(v') \nonumber \\
\rho_c(w)&=&
\int_{-v_0}^{v_0} {dv \over 2\pi}
2\dot\theta(2w-2v) \rho_s(v) 
\label{rhotj}
\end{eqnarray}
and the spinon and holon velocities are: 
\begin{eqnarray}
2\pi v_s&=&{1\over \rho_s(v_0)}{d \epsilon_s(v)\over dv }
\Big\vert_{v_0}
\nonumber \\
2\pi v_c&=&-{1\over \rho_c(w_0)}{d \epsilon_c(w)\over d w }
\Big\vert_{w_0}
\label{velocitj}
\end{eqnarray}
The cut--off $v_0$ and the holon rapidity $w_0$ are
related to the magnetization and to the position of the hole in the 
charge distribution by
\begin{eqnarray}
{I_h\over L}&=&\int_{-v_0}^{v_0} dv \rho_s(v)\theta(2w_0-2v)\nonumber\\
{1+N_s\over L}&=&\int_{-v_0}^{v_0} dv \rho_s(v)
\label{magtj}
\end{eqnarray}
Other relevant quantities are the dressed charges:
\begin{eqnarray}
\xi_{22}(v) &=&1
- \int_{-v_0}^{v_0} {dv'\over 2\pi} 
\dot\theta(v-v') \xi_{22}(v') \nonumber \\
\xi_{12}(w)&=&
\int_{-v_0}^{v_0} {dv\over 2\pi} 
2\dot\theta(2w-2v) \xi_{22}(v) 
\label{chargestj}
\end{eqnarray}
In the following, we will consider the dressed charges evaluated at the cut-off:
$\xi_{22}\equiv \xi_{22}(v_0)$ and $\xi_{12}\equiv\xi_{12}(w_0)$.
Finally, the elements of the $Z$-matrix are defined by:
\begin{eqnarray}
Z_{11}&=&{1\over 2} +
\int_{-v_0}^{v_0} {dv\over 2\pi} 
\sigma_1(v)\theta(2w_0-2v) \nonumber \\
Z_{12}&=&{1\over 2}\left [{\theta(2v_0-2w_0)-\theta(2v_0+2w_0)\over 2\pi}-
\int_{-v_0}^{v_0} {dv\over 2\pi} \sigma_1(v)
[\theta(v_0-v)-\theta(v_0+v)]\right ]
\nonumber \\
Z_{21}^+-Z_{21}^-&=& 0
\label{zetatj}
\end{eqnarray}
where the function $\sigma_1(v)$ satisfies the following equation:
\begin{equation}
\sigma_1(v)={1\over \pi}\dot\theta(2v-2w_0)-
\int_{-v_0}^{v_0} {dv'\over 2\pi} \dot\theta(v-v')\sigma_1(v')
\label{sigmatj}
\end{equation}

\vfill\eject

\vfill\eject

\centerline{\bf FIGURE CAPTIONS}
\vskip 1 true cm
\begin{itemize}
\item{\bf Fig. 1} 
Exact spectral function $A(\omega)$ (in units of the hopping $t$)
for the single hole in the Ising
model for two values of $J$. The heavy line represents the $\delta$
contribution and its height is proportional to the quasiparticle weight.

\item{\bf Fig. 2} Overlap square $Z$ between the Heisenberg ground state 
on a $\ell=L-1$ site ring and the 
ground state of the effective spin Hamiltonian
(\protect{\ref{hterm}}) corresponding to a single hole in the $L$ site $t-J$ 
model at $J=4t$ and $p=\pi/2$. 
Lanczos diagonalization has been performed 
on even chains with $L \le 26$. The dashed line is a guide to the eye.

\item{\bf Fig. 3} Overlap square $Z$ between the Heisenberg ground state on 
a $L$ site ring and the ground state of the effective spin Hamiltonian
(\protect{\ref{hterm}}) of a single hole in the $L$ site $t-J$ model at $t=0$ 
(static limit). Lanczos diagonalization has been 
performed on even chains with $L \le 26$. The dashed line is a parabolic fit
of Lanczos data. The full line shows the expected asymptotic
slope of the curve 
on the basis of conformal field theory ($2X_0=3/8$).

\item{\bf Fig. 4} Leading exponent $X_0$ of the size dependence of the overlap
$\zeta$ in the $t-J_{XY}$ model as a function of the number $N$ of 
up spins. Lines represent the numerical evaluation
of the determinant of the matrix (\protect{\ref{mij}}). Dots are the analytical 
prediction based on Eq. (\protect{\ref{phashiftxy}}).

\item{\bf Fig. 5} 
(a) Spinon term $Z(Q)$ for the $t-J_{XY}$ model computed as the
determinant of the matrix (\protect{\ref{overmat}}) at $\tau=0$ for 
magnetization $\mu=0$ (squares) and $\mu=1/4$ (triangles). Two cases 
corresponding to about $400$ and $800$ particles are shown by open and 
full symbols respectively. 
(b) Logarithmic plot of $Z(Q)$ near the singularity. The lines indicate the 
slope predicted on the basis of the finite size corrections to the energy
(\protect{\ref{phashiftxy}}).

\item{\bf Fig. 6} Asymptotic behavior of the dynamical quasiparticle weight 
$Z(R,\tau)$ in the $t-J_{XY}$ model at $\mu=-1/4$
calculated as the determinant of the matrix (\protect{\ref{brs}}) in a system 
of $201$ particles. $Z_0(R,\tau)$ is defined in Eq. (\protect{\ref{detb}}). 
Data refer to more than a thousand different points in the $(R,\tau)$ plane.

\item{\bf Fig. 7} Critical exponents $X_0$ (full squares) and $X_{-1}$ (open   
squares) as a function of the momentum of the hole $p$ for several
magnetizations $\mu=-0.4,-0.3,-0.2,-0.1,0.1,0.2,0.3,0.4$ 
from (a) to (h) respectively, in the Hubbard model at $U=4t$.

\item{\bf Fig. 8} Loci of the singularities of the spectral function in the 
$(p,\omega)$ plane (\protect{\ref{locus}}) for the Hubbard model at $U=4t$. 
Energies are measured from the bottom of the holon band. Two magnetizations
are shown: $\mu=0$ and $\mu=1/4$. Full (dashed) lines correspond to critical
exponents $X$ smaller (larger) than $1/2$. In the latter case,
according to Eq. (\protect{\ref{singa}}), the divergence disappears.

\item{\bf Fig. 9} 
Leading exponent $X_m$ of the finite size scaling of the 
overlap $\zeta$  defined in Eq. (\protect\ref{zetadef}) as a function
of the hole momentum $p$. Data refer to the $t-J$ model at $\mu=\pm 1/4$ 
and $J=2t$ in panel (a) and $J=t$ in panel (b). 
Solid lines: analytical results obtained from the finite size corrections to 
the energy in the Bethe ansatz solution.
Full (open) dots: Power law fit of Lanczos data for $\mu=1/4$ ($\mu=-1/4$).
Numerical data (full squares) 
obtained with the same fit for the non integrable point $J=t$,
are  shown for comparison only in the $\mu=1/4$ case.
The dashed line connecting these points  is a guide to the eye. 
Diagonalizations have been performed on rings with $L=8,\,16,\,24,\,32$.

\item{\bf Fig. 10} Breakdown of the completeness condition 
(\protect{\ref{completa}}) in the single spinon approximation for the 
$t-J_{XY}$ model at two magnetizations. $L$ is the length of the chain.

\item{\bf Fig. 11} Spinon function $Z(Q)$ for the Hubbard model at $\mu=0$
according to the single spinon approximation.
(a) refers to $U=4t$ and $p=p_F$; (b) to $U=8t$ and $p=p_F$;
(c) to $U=4t$ and hole momentum $p=0$; (d) to $U=8t$ and $p=0$.
The hole Fermi
momentum $p_F=\pi/2$ corresponds to the bottom of the (down) hole band.
$Z(Q)$ vanishes inside the spin up Fermi sphere.
Symbols represent Lanczos data for different system sizes: $L=6$ 
(open circles), $L=10$ (open squares), $L=14$ (full circles). 

\item{\bf Fig. 12} Same as Fig. 11 at magnetization $\mu=1/4$ corresponding to
$p_F=\pi/4$. Symbols represent Lanczos data for different system sizes: $L=8$ 
(open circles), $L=12$ (open squares), $L=16$ (full circles). 

\item{\bf Fig. 13} Hole spectral function in the Hubbard model at $U=4t$
as a function of frequency measured from the bottom of the holon band.
Energy is in units of the hopping $t$ and the calculation has been
performed in single spinon approximation. 
Panel (a) corresponds to $p=0$ and $\mu=0$, (b) to $p=0$ and $\mu=1/4$, 
(c) to $p=p_F$ and $\mu=0$, (d) to $p=p_F$ and $\mu=1/4$. 
Vertical lines show the location of
singularities. Dashed lines identify the divergences induced by
band structure effects; dotted lines show the frequencies of the
non trivial singularities reported in Fig. 8.

\end{itemize}


\begin{thebibliography}{99}
\bibitem{foot}
An analogously surprising performance of the single
spinon approximation has been already noticed in the calculation
of the Green function for the Calogero Sutherland model \cite{ha}
where the one particle intermediate states provide an excellent approximation
to the exact result.
\end{thebibliography}
\end{document}